\documentclass[sn-nature]{sn-jnl}

\usepackage{graphicx}%
\usepackage{multirow}%
\usepackage{amsmath,amssymb,amsfonts}%
\usepackage{amsthm}%
\usepackage{mathrsfs}%
\usepackage[title]{appendix}%
\usepackage{xcolor}%
\usepackage{textcomp}%
\usepackage{manyfoot}%
\usepackage{booktabs}%
\usepackage{algorithm}%
\usepackage{algorithmicx}%
\usepackage{algpseudocode}%
\usepackage{listings}%
\usepackage{tabularx}%

\theoremstyle{thmstyleone}%
\theoremstyle{thmstyletwo}%

\theoremstyle{thmstylethree}%

\begin{document}

\title{Shaping Passenger Experience: An Eye-Tracking Study of Public Transportation Built Environment}


\author[1]{\fnm{Yasaman} \sur{Hakiminejad}}\email{yhakimin@villanova.edu}

\author[2]{\fnm{Elizabeth} \sur{Pantesco}}\email{Elizabeth.Pantesco@villanova.edu}

\author*[3]{\fnm{Arash} \sur{Tavakoli}}\email{arash.tavakoli@villanova.edu}

\affil[1]{\orgdiv{Civil and Environmental Engineering}, \orgname{Villanova University},\orgaddress{\street{800 E Lancaster Ave}, \city{Villanova}, \postcode{19085}, \state{PA}, \country{United States}}}

\affil[2]{\orgdiv{Department of Psychological and Brain Sciences}, \orgname{Villanova University}, \orgaddress{\street{800 E Lancaster Ave}, \city{Villanova}, \postcode{19085}, \state{PA}, \country{United States}}}

\affil*[3]{\orgdiv{Civil and Environmental Engineering}, \orgname{Villanova University},\orgaddress{\street{800 E Lancaster Ave}, \city{Villanova}, \postcode{19085}, \state{PA}, \country{United States}}}


\abstract{Designing public transportation cabins that effectively engage passengers and encourage more sustainable mobility options requires a deep understanding of how users from different backgrounds, visually interact with these environments. The following study employs eye-tracking technology to investigate visual attention patterns across six distinct cabin designs, ranging from the current and poorly maintained versions to enhanced, biophilic focused, cyclist-friendly, and productivity-focused configurations. A total of N=304 participants engaged with each cabin design while their eye movements such as Fixation Counts, Time to First Fixation (TFF), First Fixation Duration (FFD), Stationary Gaze Entropy (SGE), and Gaze Transition Entropy (GTE) were recorded. Results revealed that alternative cabin configurations consistently exhibited shorter TFFs and lower entropy measures compared to the baseline current version. Specifically, designs incorporating natural elements and biophilic aspects, streamlined layouts, or functional amenities, facilitated quicker orientation and more structured gaze patterns, indicating enhanced visual engagement and possibly reduced cognitive load. In contrast, the poorly maintained cabin design was associated with higher entropy values, suggesting more scattered and less predictable visual exploration. Demographic factors, particularly ethnicity, significantly influenced FFD in certain designs, with Non-white participants showing reduced fixation durations in the enhanced and poorly maintained environments highlighting the importance of inclusive design considerations. Moreover, transportation-related demographic factors—such as frequency of public transport use, trip purpose, and duration of use—significantly influenced visual attention metrics in various cabin designs. Participants with lower frequencies of public transport use oriented their gaze more rapidly in visually enhanced environments, implying that thoughtful design can attract and satisfy infrequent riders. By leveraging eye-tracking insights, designers and policymakers can make evidence-based modifications that not only enhance the aesthetic and functional quality of public transport cabins but also promote broader transit adoption and support sustainability objectives.

}

\keywords{Public transportation cabin design, Eye Tracking, Gaze Entropy, Fixation, Biophilic transportation spaces}



\maketitle

\section{Introduction}
Public transportation is an integral component of urban infrastructure, facilitating the mobility of millions of people daily and contributing to economic growth and environmental sustainability \cite{zhou2023exploring,chatman2011public,nanaki2017environmental}. As cities around the world face challenges such as traffic congestion, pollution, and the need for efficient and sustainable resource utilization; the role of trains, buses, and other public transit systems becomes increasingly vital. 

Enhancing public transportation to support sustainable transportation moves beyond operational efficiency and environmental benefits to focus on creating systems that are inclusive, adaptive, and appealing to diverse user needs \cite{nahiduzzaman2021influence,te2019potential}. An important aspect of such systems is the \textbf{ultimate user experience}. Enhancing the user experience within these systems is paramount not only for maintaining current ridership but also for encouraging a shift away from private vehicle use, thus reducing urban carbon footprints \cite{mcleod2017urban,adewumi2013rea,foth2010enhancing}. Previous research shows that the physical environment of public transport cabins significantly influences passengers' comfort, perception of safety, and overall willingness to use these services regularly \cite{morton2016customer,friman2020public,ingvardson2022influence}. Thus, there is a growing need to rethink and redesign these spaces to create more welcoming, functional, and user-centered environments \cite{putri2024ergonomic,hakiminejad2024public,stolz2021user}.

Redesigning public transportation cabins involves more than just aesthetic enhancements; it requires a comprehensive understanding of how different design elements affect the passenger experience \cite{hakiminejad2024public,tyrinopoulos2008public}. Similar to any other built environments \cite{hollander2019seeing}, one major aspect of this relationship is understanding how passengers visually interact with different cabin designs and how demographic factors influence these interactions. Traditional methods of assessing passenger experience often rely on surveys and self-reported measures, which can be subject to biases and may not capture the nuances of real-time user engagement\cite{zammarchia2023using,schrammel2011attentional,piao2013capturing,noland2017eye}. To address this gap, there is growing interest in using objective metrics such as \textbf{eye tracking} technology for measuring visual attention and cognitive processing in response to environmental variations in public transportation environments.

Eye tracking offers a unique window into passengers' unconscious preferences and attentional priorities by recording where, when, and how long individuals look at specific elements within a cabin space \cite{shiferaw2019review,holmqvist2011eye,hollander2019seeing}. Metrics such as fixation duration, fixation locations and entropy measures provide quantifiable data on visual engagement, which can inform designers about which features attract attention and how users visually navigate the environment \cite{shiferaw2019review,holmqvist2011eye,hollander2019seeing}. Understanding how various groups of users from different demographics (e.g., ethnicity, gender, public transportation use) interact with the new designs can result in efficient and more human-centric design choices, which will ultimately lead to an increase in public transportation use.

In this study, we aim to explore the intersection of public transportation cabin design, visual attention, and demographic differences by employing eye-tracking technology on top of regular surveying techniques. We presented participants with six distinct cabin interior designs:\textit{Current Version}, \textit{Poorly Maintained Current Version}, \textit{Enhanced Version}, \textit{Biophilic Design}, \textit{Bike-Centered Design}, and \textit{Productivity-Focused Design}. These designs represent a spectrum of existing and potential future cabin configurations, each with unique features intended to address specific passenger needs and preferences.

Our objectives are threefold: First, to determine whether different cabin designs elicit varying patterns of visual attention among passengers. Second, to assess how demographic factors such as frequency of public transport use, purpose of trip, and duration of use influence these patterns. Third, to provide actionable insights that can guide the development of more effective and user-friendly public transportation cabin designs. By integrating eye-tracking metrics with demographic analysis, this research contributes to a more nuanced understanding of passenger experience in public transit environments. By shedding light on how passengers of diverse backgrounds visually perceive and interact with various cabin environments, this study aims to support the creation of public transportation spaces that are not only efficient but also engaging and responsive to the needs of all users. This information is crucial for transportation authorities, designers, and policymakers aiming to enhance the appeal and functionality of public transit systems.

The remainder of this paper is structured as follows: we first provide a background on eye-tracking methodologies and their relevance to environmental design and public transportation research. We will then detail our methodology, including participant recruitment, experimental procedures, and the specific eye-tracking metrics employed. The next section presents the results of our analysis, highlighting significant findings related to fixation counts, time to first fixation, fixation durations, and entropy measures across different cabin designs and demographic groups. We will then discuss the implications of our findings, addressing how they inform design considerations and the potential limitations of our study. Finally, we conclude by making recommendations for future research directions in the field of public transportation design.





\section{Background}

\subsection{User Experience and the Built Environment}

\begin{figure}[ht]
    \centering
    \includegraphics[width=0.6\linewidth]{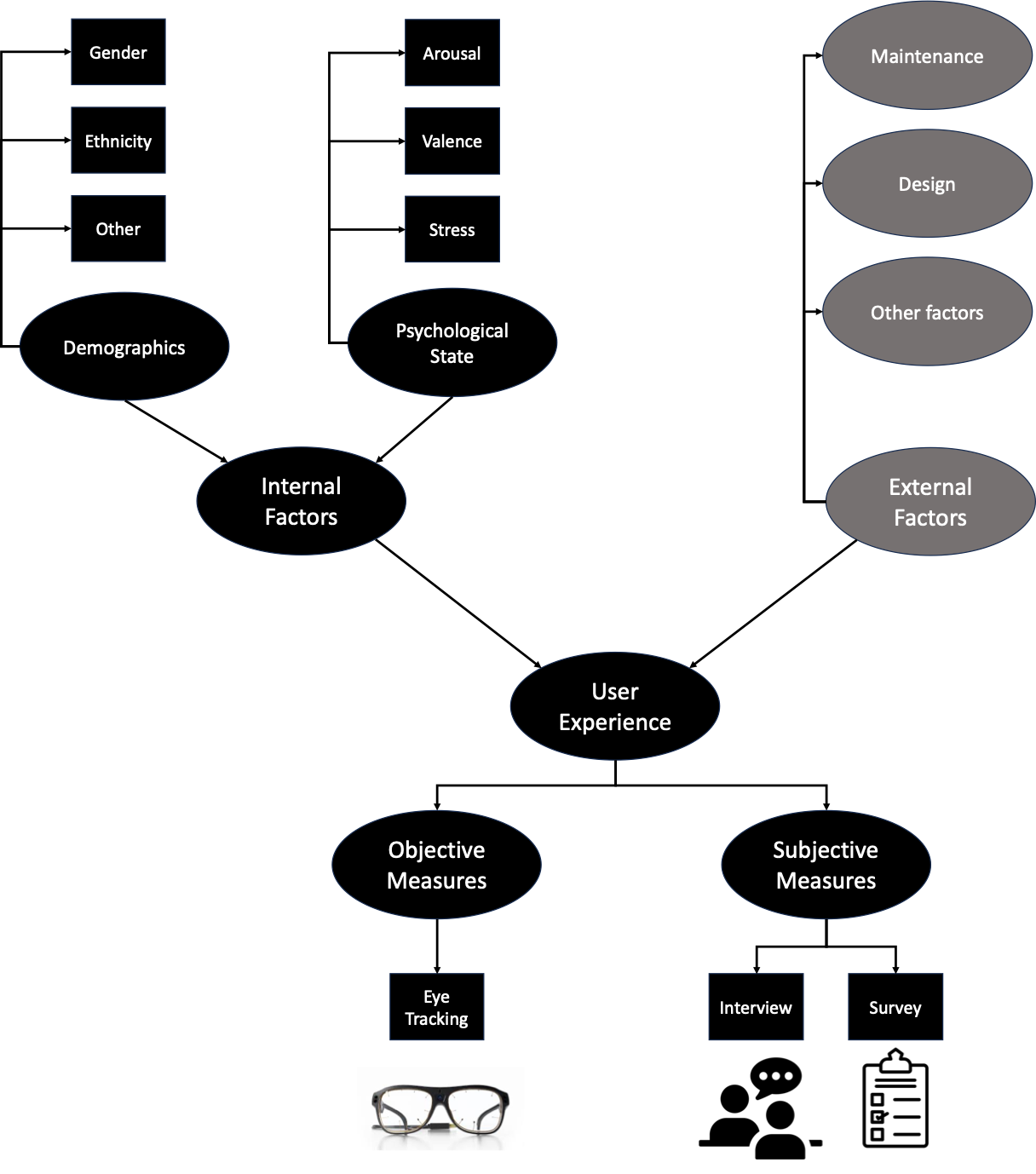} 
    \caption{Overview of the various aspects affecting user experience in public transportation}
    \label{fig:background}
\end{figure}

User experience in public transportation is a multifaceted construct, encompassing physical comfort, perceived safety, reliability, convenience, and the emotional response elicited by perceiving the transit environment \cite{redman2013,carrel2013,abou2012happiness}. Consistent with prior research on the built environment \cite{tavakoli2022multimodal,tavakoli2025psycho,guo2023psycho,nussbaumer2018human,ibrahim2021mood,honghumanising}, user experience is impacted by both internal and external factors, as shown in Fig. \ref{fig:background}.

Internal factors encompass demographic characteristics and user-related states, which significantly influence how individuals perceive and interact with public transportation environments \cite{honghumanising}. Demographic characteristics such as age, gender, income level, and cultural background play a crucial role in shaping expectations and preferences related to public transportation \cite{cantwell2009examining,susilo2014exploring}. Additionally, transportation related demographics also play an internal role. For instance, infrequent or first-time users may struggle to orient themselves in unfamiliar cabin layouts, whereas experienced commuters can more easily navigate complex environments \cite{Beirao2007, Anable2005}. Similarly, passengers who predominantly use public transit for work-related travel may focus on efficiency and reliability, while leisure travelers may pay more attention to aesthetic quality and comfort \cite{VanLierop2016}. Understanding these diverse user profiles is essential for developing design solutions that meet the needs of a heterogeneous ridership \cite{susilo2014exploring} and move toward human-centric public transportation systems. Tailoring cabin features and informational materials to various user segments can enhance the overall travel experience, potentially improving satisfaction and loyalty towards public transportation systems.

Another aspect of internal factors includes user-related states, such as mood, fatigue, and cognitive load, which significantly influence the perception of transit environments \cite{grison2016exploring,pedersen2011affective}. For instance, commuters experiencing stress or fatigue prior entering public transit may perceive the same environment as more overwhelming compared to those in a relaxed state, affecting their satisfaction and engagement with the system. Cognitive load, often heightened by complex navigation or unclear signage, can further hinder the ability to interpret and respond to environmental cues, especially for first-time users. Addressing these states through intuitive design and clear communication can reduce cognitive strain and improve overall user experience \cite{pedersen2011affective}.

On the other hand, external factors include various physical features and maintenance-related aspects that can significantly influence user experience in public transportation \cite{Abdelkarim2023}. Studies show that elements such as cabin cleanliness, ergonomic seating, proper ventilation, and lighting play a critical role in shaping how users perceive the quality and comfort of transit systems \cite{IMRE20172441,redman2013,carrel2013}. For example, well-maintained interiors with clean surfaces and functional amenities not only create a positive impression but also enhance users' sense of safety and reliability \cite{redman2013, carrel2013}. Additionally, the presence of aesthetic features, such as greenery, artwork, or thoughtfully designed spaces, similar to other built environments \cite{altaf2023time,bianchi2023study,douglas2022physical} can evoke positive emotional responses and reduce stress levels during transit \cite{hakiminejad2024public}. Maintenance-related factors, such as timely repairs, graffiti removal, and consistent upkeep, reinforce trust in the system by signaling professionalism and care for passengers \cite{Eboli2014}. Accessibility features, such as ramps, elevators, and clear signage, are also critical external factors that affect user experience \cite{kapsalis2024disabled}. These features ensure that the system is inclusive and functional for all users, including those with mobility challenges or other disabilities. Moreover, the integration of noise control measures and temperature regulation within cabins can further enhance the overall experience by addressing comfort and environmental stressors \cite{kryter2010handbook, Babisch2011}.

\subsection{Quantifying User Experience}

To better understand user interaction with the transportation built environment, various methodologies have been used in the past, where many of these methods are subjective and prone to biases. For instance, traditional surveys and interviews rely heavily on self-reported data, which can be influenced by memory recall errors, social desirability, or individual perceptions that may not accurately reflect real-time experiences \cite{rosenman2011measuring}. These limitations highlight the need for objective, data-driven methods to gain a deeper understanding of user interactions and responses.

Among the objective approaches, eye-tracking technology has emerged as a powerful tool for analyzing user experience \cite{tavakoli2022multimodal,guo2023psycho,tavakoli2025psycho,shiferaw2019review}. Eye tracking is a well-established methodology in human factors, cognitive psychology, and human-computer interaction research, providing objective evidence of where, when, and how long individuals direct their visual attention \cite{shiferaw2019review}. By recording eye movements such as saccades, fixations, and scanning paths, researchers gain insight into cognitive processing, user preferences, and task performance in a range of contexts across engineering, psychology, and health \cite{Duchowski2017, JacobKarn2003,pashkevich2021operation,guo2023rethinking,guo2023psycho,tavakoli2022multimodal,krejtz2015gaze}. In recent years, eye-tracking technology has become more accessible, with mobile and remote trackers enabling studies in more ecologically valid environments and by enhancing the reach to a wider audience \cite{tavakoli2025psycho,wisiecka2022comparison,murali2021conducting,kyrk2023eye,heck2023webcam}. 

Within the domain of transportation research, eye tracking has proven valuable for understanding various aspects of driver behavior \cite{tavakoli2022multimodal}, pedestrian and cyclist behavior \cite{guo2023psycho,guo2023rethinking}, as well as how users interact with the built environment \cite{hollander2019seeing,tavakoli2025psycho}. Extending this methodology to public transportation cabins allows researchers to understand how passengers visually navigate interior layouts, interpret signage, and identify amenities, thereby revealing opportunities to enhance comfort and usability through design interventions \cite{holler2009perception}.

\subsection{Eye Tracking Data: Feature Extraction and Analysis}

To analyze eye-tracking data, several features are derived from the raw gaze data \cite{shiferaw2019review,tavakoli2022multimodal,tavakoli2025psycho}. The first key metric is fixation points, which represent moments when the gaze remains stationary on a particular object or area. In contrast, a saccadic movement refers to the rapid shifts of the gaze between points of fixation. From fixation data, several features can be calculated, including time to first fixation (TFF), first fixation duration (FFD), the total number of fixation points, the average fixation duration, stationary gaze entropy, and gaze transition entropy \cite{tavakoli2022multimodal,shiferaw2019review,guo2023psycho,guo2023rethinking}.

First Fixation Duration (FFD) and Time to First Fixation (TFF) provide insights into visual salience and attention \cite{shiferaw2019review}. FFD measures the duration of the initial fixation on a specific visual element, indicating its perceived importance or ease of processing. Shorter durations may suggest familiarity and the need to explore other stimuli \cite{nahari2024fixation} or immediate recognition. TFF measures the time it takes for the gaze to reach a specific visual element after it is presented, highlighting the prominence and salience of the element.

In addition to linear metrics, this study employs two entropy-based eye-tracking metrics, Stationary Gaze Entropy (SGE) and Gaze Transition Entropy (GTE), to comprehensively assess visual engagement and attention patterns within various public transportation cabin designs \cite{tavakoli2022multimodal,guo2023psycho,shiferaw2019review}. SGE quantifies the randomness in the spatial distribution of fixations across different cabin elements, while GTE measures the unpredictability of gaze transitions between these elements. Together, these metrics provide deeper insights into participants' visual exploration strategies and the intensity of their attention. This dual assessment enables the identification of whether passengers engage with the cabin environment in a deliberate and concentrated manner or adopt a more random and exploratory approach, thereby informing design modifications that enhance passenger experience by promoting desired visual engagement behaviors.

More specifically, SGE quantifies the unpredictability in the spatial distribution of a participant's fixations across predefined Areas of Interest (AOIs) within a visual field. It serves as an indicator of how dispersed or concentrated a participant's visual attention is during the observation of complex environments. Mathematically, let \( \mathcal{A} = \{A_1, A_2, \dots, A_n\} \) represent a set of \( n \) non-overlapping AOIs within the cabin design. During a viewing session, let \( f_i \) denote the number of fixations made within AOI \( A_i \). The probability \( p_i \) of a fixation occurring in AOI \( A_i \) is calculated as:

\[
p_i = \frac{f_i}{F}
\]

where \( F = \sum_{j=1}^{n} f_j \) is the total number of fixations across all AOIs.

The Stationary Gaze Entropy (SGE) is then defined using Shannon's entropy formula:

\[
\text{SGE} = -\sum_{i=1}^{n} p_i \log_2(p_i)
\]

A high SGE value indicates a broad and uniform distribution of fixations across multiple AOIs, reflecting exploratory viewing behavior and a wide spread of attention. Conversely, a low SGE value suggests that fixations are concentrated in a few AOIs, indicating focused and deliberate attention towards specific visual elements \cite{shiferaw2019review}.

On the other hand, GTE measures the unpredictability of transitions between different AOIs during a viewing session. It captures the complexity and randomness in the sequence of visual attention shifts, thereby reflecting whether gaze movements follow a systematic pattern or are random and exploratory. Consider the same set of AOIs \( \mathcal{A} = \{A_1, A_2, \dots, A_n\} \). Let \( T_{ij} \) denote the probability of transitioning from AOI \( A_i \) to AOI \( A_j \). This transition probability is estimated as:

\[
T_{ij} = \frac{t_{ij}}{\sum_{k=1}^{n} t_{ik}}
\]

where \( t_{ij} \) represents the number of observed transitions from AOI \( A_i \) to AOI \( A_j \).

Gaze Transition Entropy (GTE) is then defined as the conditional entropy of the next fixation given the current fixation:

\[
\text{GTE} = -\sum_{i=1}^{n} p_i \sum_{j=1}^{n} T_{ij} \log_2(T_{ij})
\]

where \( p_i \) is the stationary probability of AOI \( A_i \), calculated as:

\[
p_i = \frac{\sum_{k=1}^{n} t_{ki}}{\sum_{k=1}^{n} \sum_{m=1}^{n} t_{km}}
\]

A high GTE value implies a high level of unpredictability and randomness in gaze transitions between AOIs, indicative of exploratory and less structured viewing behavior. In contrast, a low GTE value suggests predictable and systematic transitions between AOIs, reflecting focused and goal-directed visual attention \cite{tavakoli2022multimodal}.

\section{Hypotheses}

This study posits that internal and external factors significantly influence passengers' visual engagement and overall unconscious experience. More specifically, we hypothesize that: 

\textbf{H1:} \textit{Various external factors measured through different public transportation cabin designs significantly affect visual attention metrics, including Time to First Fixation (TFF), Fixation Counts, and Gaze Entropy.}

\textbf{H2:} \textit{Internal factors such as demographic factors (i.e., ethnicity, gender, frequency of public transport use and trip purpose) as well as participants prior internal state (i.e., stress, and emotion) affect the relationship between cabin design and visual attention metrics.}

\section{Methodology}
\begin{figure}[ht]
    \centering
    \includegraphics[width=1\linewidth]{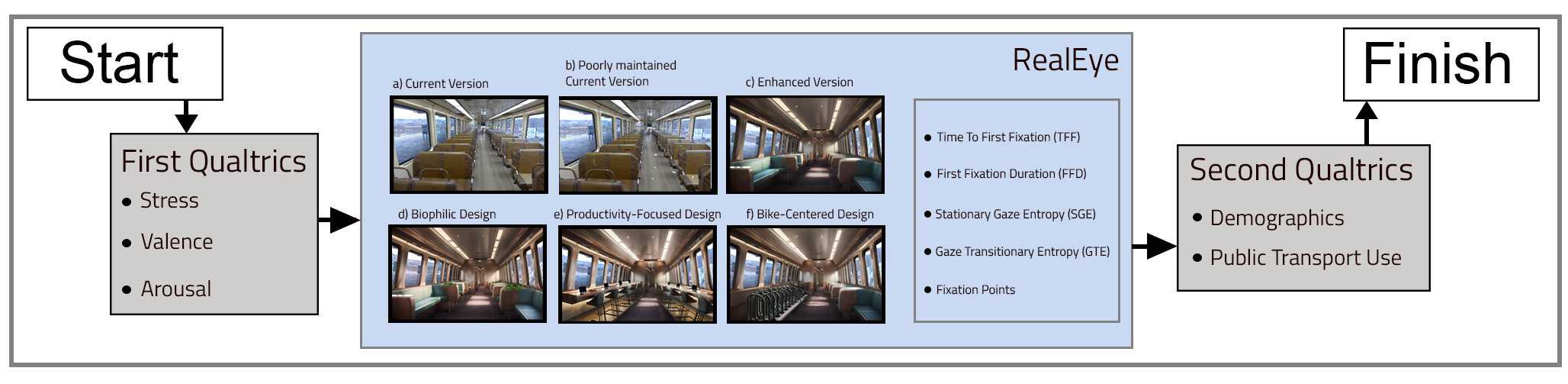} 
    \caption{The study design framework is illustrated, with blue sections indicating the RealEye platform and gray sections representing Qualtrics. The images were presented in a randomized order within RealEye.}
    \label{fig:procedure}
\end{figure}

This section outlines the methodological framework of the study, including the \textit{Study Design}, which details the within-subject approach and experimental setup; \textit{Participant Recruitment and Demographics}, which describes the recruitment process and participant characteristics; \textit{Eye-Tracking Data Analysis}, focusing on the metrics and areas of interest analyzed; and \textit{Model Development}, which explains the use of linear mixed-effects models to investigate the relationships between demographic variables, cabin designs, and visual attention patterns. A schematic view of the methodology can be viewed on Fig. \ref{fig:procedure}. The data for this study is gathered through a previous experimental framework performed by Hakiminejad et al \cite{hakiminejad2024public}. Below we provide a summary of the key information required for the subsequent analysis. 

\subsection{Study Design}

This study employs a within-subjects experimental design to investigate how demographic differences influence participants' perceptions of various public transportation cabin designs, as assessed through eye-tracking data. By utilizing a within-subjects approach, the study controls for individual differences in visual attention, ensuring that observed effects can be more confidently attributed to the design variations of the public transportation cabins.

Participants initially completed a Qualtrics survey to assess their baseline emotional and cognitive states, including momentary stress, valence (emotional positivity/negativity), and arousal (level of emotional alertness)\cite{Karvounides2016-ja, Mauss2009-sr}. These measure serve as their internal state prior to the experiment. More specifically, each participant responded to a one-item questionnaire for each metric of internal state \cite{hakiminejad2024public}. Subsequently, they were redirected to RealEye.com \cite{RealEye}. 

RealEye has been employed in numerous studies for capturing remote eye-tracking data effectively \cite{RealEye,wisiecka2022comparison,murali2021conducting,tavakoli2025psycho}. This platform utilizes an innovative webcam-based eye-tracking method that predicts a user's gaze point by leveraging the computing power of the host computer to run a deep neural networks analyzing images resulting from the computer webcam. This AI-driven process detects the participant's face and pupils, predicting gaze points in real-time. RealEye has demonstrated an accuracy of approximately 124 pixels on desktop and laptop devices and around 60 pixels on mobile devices, allowing for precise analysis of user interactions down to the size of a single button. Additionally, gaze points are predicted at frequencies of up to 60 Hz, facilitating detailed temporal analysis of user behavior \cite{RealEye}.

Within the platform, participants were first guided through a calibration process to ensure accurate eye-tracking data collection. Then the participants viewed six distinct images representing different public transportation cabin designs: \textit{Current Version}, \textit{Poorly Maintained Current Version}, \textit{Enhanced Version}, \textit{Biophilic Design}, \textit{Bike-Centered Design}, and \textit{Productivity-Focused Design}. Each image was displayed for a fixed duration of 10 seconds, during which participants' eye movements were continuously recorded through their webcam and by employing computer vision techniques performed by RealEye.

\begin{figure}[ht]
    \centering
    \includegraphics[width=1\linewidth]{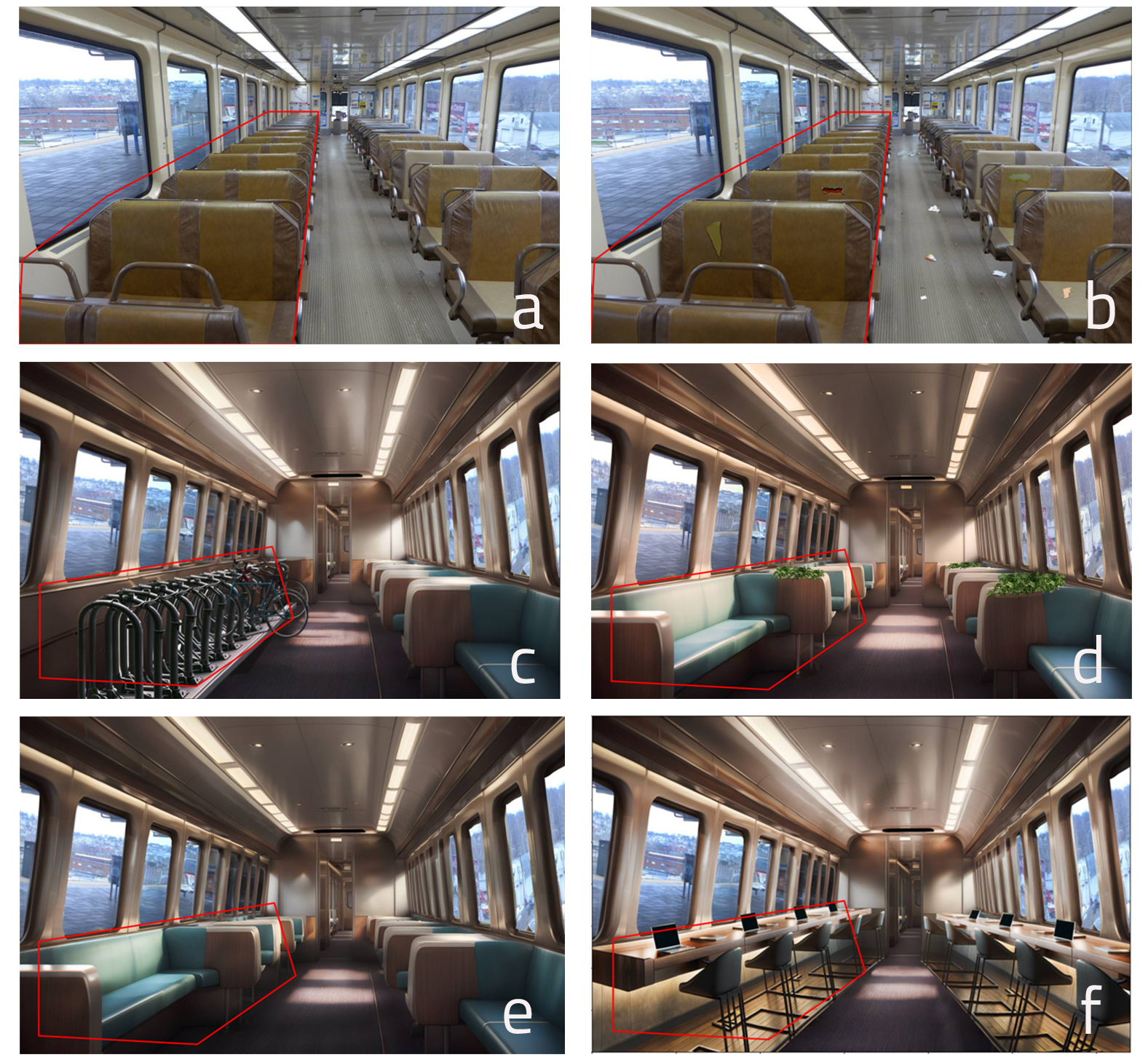} 
    \caption{Visual comparison of cabin designs used in the study, illustrating a variety of interior layouts as well as the Area of Interest Boundary in different cabin designs;
    a: \textit{Current Version}, b: \textit{Poorly Maintained Current Version}, c:\textit{Bike-Centered Design}, d: \textit{Biophilic Design}, e: \textit{Enhanced Version}, f: \textit{Productivity-Focused Design}}
    \label{fig:boundary}
\end{figure}

The cabin designs were performed using extensive graphic editing and rendering, along with utilizing advanced Artificial Intelligence (AI) technologies, including GPT-4 (DALL-E), as shown on Fig. \ref{fig:boundary}. To ensure designs stayed consistent, the main stimuli were placed on the left side of the images where applicable (Fig. \ref{fig:boundary}). Each cabin design was characterized by specific features intended to (1) address different passenger needs and preferences as well as (2) simulating cabin characteristics that can affect users such as lack of maintenance. The \textit{Current Version} depicted the standard, existing design with traditional seating arrangements and basic amenities retrieved from a Northeastern regional train \cite{hakiminejad2024public}. The \textit{Poorly Maintained Current Version} illustrated the same design in a state of disrepair, featuring worn-out seats, inadequate lighting, and cluttered spaces. The \textit{Enhanced Version} showcased improvements such as upgraded materials, ergonomic seating, advanced lighting systems, and streamlined signage. The \textit{Biophilic Design} integrated natural elements, including plants and natural light simulations, to create a more calming and aesthetically pleasing environment. The \textit{Bike-Centered Design} included dedicated bike storage areas, while the \textit{Productivity-Focused Design} incorporated work-friendly amenities such as USB charging ports, foldable tables, and partitions for semi-private workspaces. 


After viewing each image, participants answered questions regarding their willingness to use the depicted design and provided feedback on other metrics related to the environment. Upon completing the eye-tracking session, participants were directed back to Qualtrics to complete a second survey, which included a comprehensive demographic questionnaire. This questionnaire collected data on age, gender, education level, frequency of public transport use, trip purpose, duration of use, personality, psychological well-being, housing condition, and zip code of housing location.

\subsection{Participant Recruitment and Demographics}

\begin{figure}[ht]
    \centering
    \includegraphics[width=0.9\linewidth]{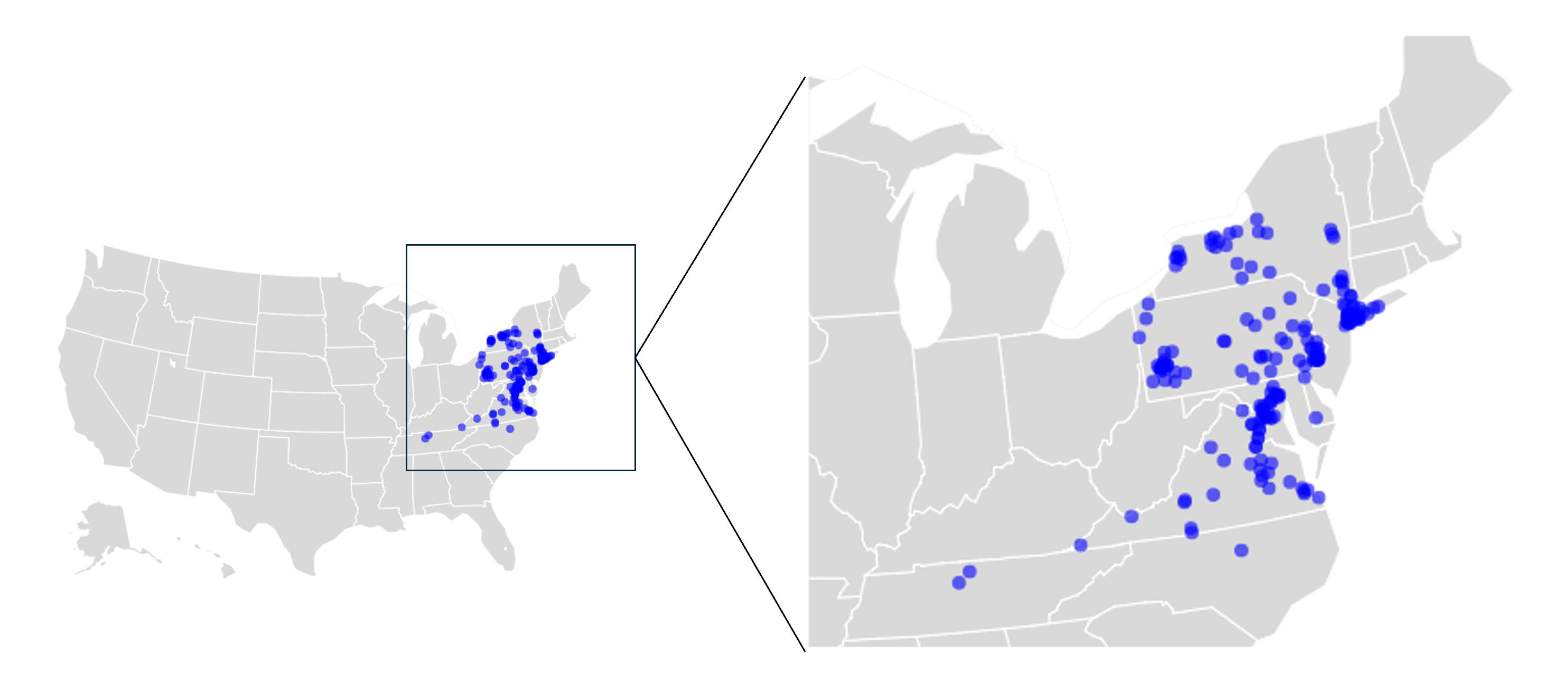} 
    \caption{Participant spread across the US}
    \label{fig:map}
\end{figure}

The study was approved by the Villanova University Institutional Review Board (IRB). 304 participants from the Northeastern region of the United States were recruited through the Prolific Platform \cite{prolific} as shown on Fig. \ref{fig:map}. Each participant received \$5.20 as compensation for their 15-minute participation in the study. The demographics of the participants are shown in Table \ref{demographics}.

\begin{table}[ht!]
\centering
\caption{Demographics of participants (N=304)}
\label{demographics}
\begin{tabularx}{\textwidth}{@{}Xccc@{}}
\toprule
\textbf{Variable} & \textbf{Attribute} & \textbf{Number} & \textbf{Percent (\%)} \\
\midrule
\textbf{Gender}  & Male & 155 & 50.98 \\
& Non-male & 149 & 49.01 \\
\midrule
\textbf{Ethnicity} & White & 213 & 70.06 \\
& Non-white & 91 & 29.93 \\
\midrule
\textbf{Education} & Less than a Bachelor’s degree & 80 & 26.31 \\
 & Bachelor’s degree or higher & 223 & 73.35 \\
 & Prefer not to say & 1 & 0.33 \\
\midrule
\textbf{Age} & Mean (Std. Dev) & 39.60 (13.45) & -- \\
 & Range & 19--78 & -- \\
\bottomrule
\end{tabularx}
\end{table}

\subsection{Eye Tracking Data Analysis}

Upon data collection, the eye tracking data was analyzed to capture various gaze metrics as follows. Outputs of the RealEye included location and duration of fixation points. RealEye provided a quality metric for the data based on participant's camera and background condition. The quality grade ranged from 1 to 6. Those with a participant quality grade below 3 were excluded, leaving 273 participants for the analysis. This can be due to low lighting, too much movement or similar reasons that degrade the eye tracking data.

Based on the fixation dataframe retrieved from RealEye, a series of features were calculated for each participant within each cabin condition. Key metrics analyzed included number of fixation points, time to first fixation, first fixation duration, gaze transition entropy, and stationary gaze entropy. These features are detailed out as follows and include both area of interest specific and image specific features:

\begin{enumerate}
    \item \textbf{Fixation count:} the number of fixations per participant within the area of interest. 
    \item \textbf{Time to First Fixation (TFF):} the amount of time it takes for a participant to make their first fixation within the area of interest. 
    \item \textbf{First Fixation Duration (FFD):} the amount of time each participant spends on their first fixation within the area of interest.
    \item \textbf{Stationary Gaze Entropy} a measure of how uniformly a participant's gaze is distributed across the image of interest as a measure of randomness of dispersion.
    
    \item \textbf{Gaze Transition Entropy} a measure of the randomness or predictability in the transitions between gaze points across the image of interest.

\end{enumerate}


\subsection{Model Development}

To analyze the relationship between eye-tracking metrics and participants' demographics, we employed a combination of linear and linear mixed-effects models where applicable as follows. Linear mixed-effects models (LMMs) are particularly well-suited for capturing individual variability by incorporating both fixed effects (e.g., cabin design) and random effects (e.g., participant-specific differences) \cite{baayen2008mixed,bates2015fitting}. This capability makes LMMs especially effective for analyzing hierarchical or repeated-measures data, such as eye-tracking metrics influenced by demographic factors and cabin design. By accounting for within-participant variability, this approach allows us to examine how various designs interact with eye-tracking metrics. In the context of our study, the linear mixed-effects model can be expressed as:

\[
y_{ij} = \beta_0 + \beta_1 \text{(Demographics)}_{ij} + \beta_2 \text{(Cabin Design)}_{ij} + b_{0i} + \epsilon_{ij}
\]

Where:
\begin{itemize}
    \item $y_{ij}$ represents the dependent variable, which corresponds to the Eye-Tracking metrics used in this study (e.g, Gaze Transition Entropy, Time to First Fixation, and Stationary Gaze Entropy)
    \item $\beta_0$ is the fixed intercept, indicating the baseline value of the dependent variable.
    \item $\beta_1$ represents the fixed effects of gender and ethnicity.
    \item $\beta_2$ represents the fixed effects of the cabin design images (e.g., Bike-Centered Design, Biophilic Design).
    \item $\text{(Demographics)}_{ij}$ denotes the gender and ethnicity variables for the $j$-th participant in the $i$-th group.
    \item $\text{(Cabin Design)}_{ij}$ denotes the specific cabin design image viewed by the $j$-th participant.
    \item $b_{0i}$ is the random intercept, accounting for variability between participants or groups (e.g., individual differences in baseline perceptions).
    \item $\epsilon_{ij}$ is the residual error term, assumed to follow a normal distribution $(\epsilon_{ij} \sim N(0, \sigma^2))$.
\end{itemize}

This model evaluates how demographic characteristics and the specific cabin design images influence participants' perceptions, as measured through eye-tracking metrics. By including both fixed and random effects, the model captures individual variability while examining the broader effects of demographics and cabin designs on participant perceptions. After comparing various designs based on participants perceptions, we also compared participants' eye tracking responses within each image based on their transportation related demographics through linear models to understand how various groups of participants (e.g., frequent users versus non frequent users) perceive each design.



\section{Results}
To present our findings in a structured and coherent manner, we begin by analyzing the specific locations where participants directed their attention across the various cabin designs. Subsequently, in subsection \ref{sec:eye}, we provide a comparative analysis of cabin perceptions based exclusively on eye-tracking metrics. Finally, subsection \ref{sec:demo} explores how demographic variables relate to eye-tracking measures associated with these cabin perceptions.

\subsection{Fixation Patterns Across Cabin Designs}



\begin{figure}[ht]
    \centering
    \includegraphics[width=1\linewidth]{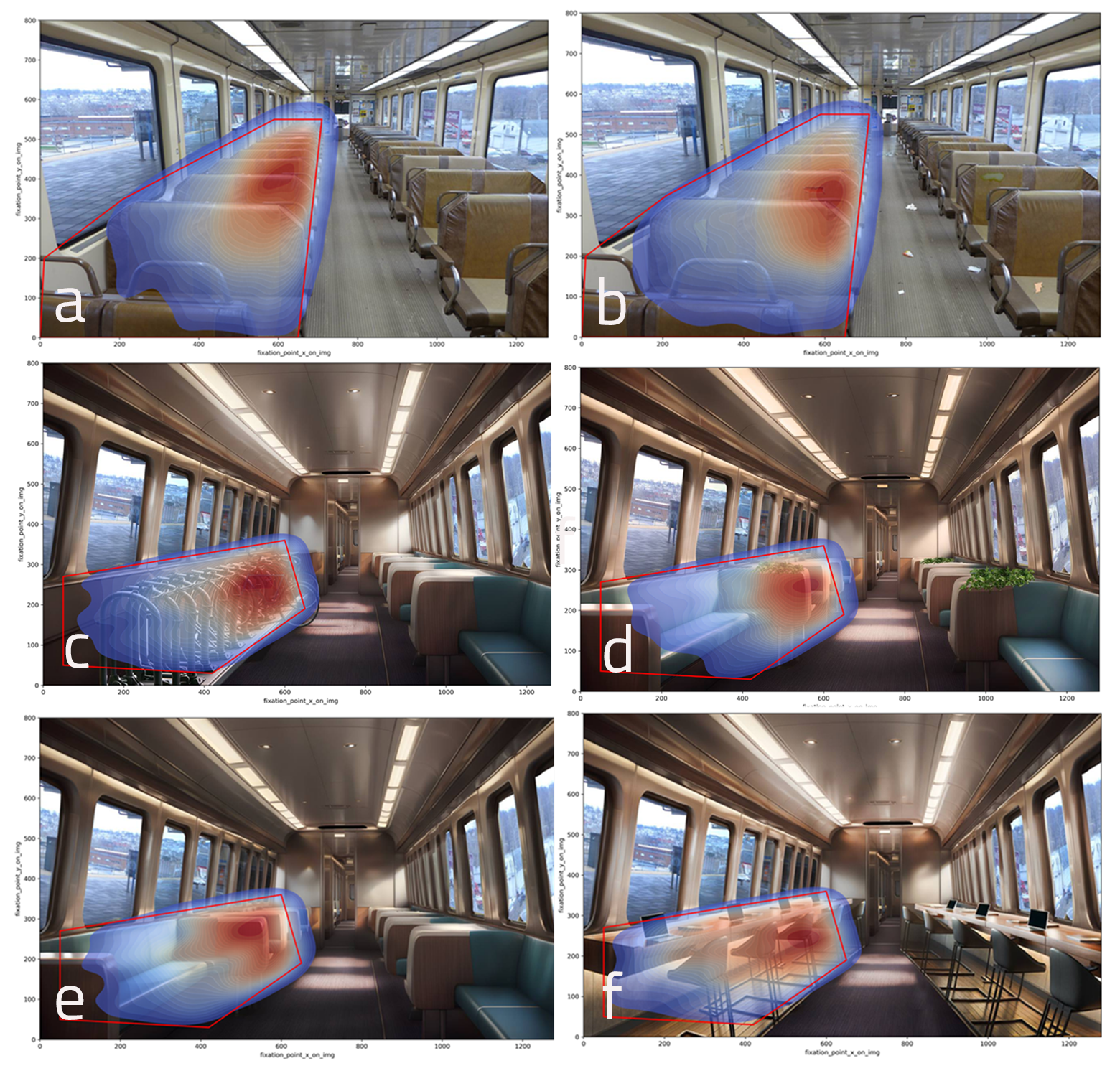} 
    \caption{Heatmaps of fixation points in the Area of Interest in different cabin design
    a: \textit{Current Version}, b: \textit{Poorly Maintained Current Version}, c:\textit{Bike-Centered Design}, d: \textit{Biophilic Design}, e: \textit{Enhanced Version}, f: \textit{Productivity-Focused Design}}
    \label{fig:heatmaps}
\end{figure}

The heatmaps presented in Figure \ref{fig:heatmaps} illustrate participants' fixation patterns across six distinct cabin designs, offering insights into areas of heightened visual attention. The red regions in the heatmaps represent high fixation density, indicating where participants focused their gaze most frequently. In the Current Version (Figure \ref{fig:heatmaps} - a), participants predominantly fixated on the seats located in the middle of the cabin—likely due to the design’s uniformity and straightforward layout. By contrast, the Poorly Maintained Current Version (Figure \ref{fig:heatmaps} - b) exhibits more dispersed attention, particularly drawn to obvious signs of wear and tear, scuff marks, and general clutter. These indicators of neglect appear to act as strong visual stimuli, highlighting how the lack of maintenance can significantly influence where viewers direct their gaze. The Bike-Centered Design (Figure \ref{fig:heatmaps} - c) significantly redirected participants’ attention to the bicycle area, with the highest fixation density concentrated on the bikes themselves. This finding underscores the bike’s salience as a focal stimulus, demonstrating the design’s effectiveness in capturing visual attention.

In the Biophilic Design (Figure \ref{fig:heatmaps} - d) participants’ gaze was drawn towards both the seating area and greenery elements. The presence of plants introduced an additional visual anchor, balancing attention between functional (seating) and aesthetic (plants) components. Meanwhile, the Enhanced Version (Figure \ref{fig:heatmaps} - e) revealed a similar concentration of fixations on seating, with a subtle shift towards the central aisle. This pattern suggests that the streamlined layout of the enhanced cabin promoted a more evenly distributed gaze, reducing visual clutter and fostering a structured viewing experience. Finally, in the Productivity-Focused Design (Figure \ref{fig:heatmaps} - f), participants’ attention shifted towards the workstations on the right side of the cabin. By integrating functional elements such as desks and chairs, the design successfully drew participants’ gaze away from the standard seating areas and toward the workspace—emphasizing productivity-related features.

\subsection{Individual Eye-Tracking Metrics}\label{sec:eye}

\subsubsection{Fixation Counts}
A comparison of fixation counts in the areas of interest between the images was carried out using linear mixed effect model. The results indicate that none of the tested designs elicited statistically significant differences in fixation counts when compared to the baseline intercept (\textit{Current version}). Among the designs, the \textit{Enhanced version} demonstrated the largest negative effect (-0.33333) and exhibited a borderline significance (\(p = 0.0926\)). However, this did not meet the standard threshold for statistical significance (\(p < 0.05\)). Similarly, other designs, including \textit{Poorly maintained current version}, \textit{Bike}, \textit{Productivity}, and \textit{Biophilic design}, exhibited minor positive or negative effects, but none achieved statistical significance (\(p > 0.05\)). These findings may suggest that participants did not exhibit substantial differentiation among the designs in terms of their fixation counts, although the possibility of differences cannot be ruled out, potentially due to limited sample size or variability in the data.






\begin{table}[ht]
\centering
\caption{Comparison of fixation counts across different cabin designs using a linear mixed effect model}
\begin{tabularx}{\textwidth}{@{}Xcccccc@{}}
\toprule
\textbf{Variable} & \textbf{Estimate} & \textbf{Std. Error} & \textbf{df} & \textbf{t value} & \textbf{Pr(>|t|)} & \textbf{Significance} \\
\midrule
(Intercept)                              & 30.540 & 0.2484 & 485.0  & 122.927  & $<2e^{-16}$ & ***  \\
Poorly maintained current version & 0.0637 & 0.2008 & 1330.0 & 0.317    & 0.7513     &      \\
Bike     & -0.3071 & 0.2008 & 1330.0 & -1.529   & 0.1264     &      \\
Productivity    & 0.0037  & 0.2008 & 1330.0 & 0.019    & 0.9851     &      \\
Biophilic design & -0.2060 & 0.2008 & 1330.0 & -1.026   & 0.3052     &      \\
Enhanced version         & -0.3483 & 0.2008 & 1330.0 & -1.734   & 0.0831     & .    \\
Gender: Non-male                       & 0.2523 & 0.2872 & 264.0  & 0.879    & 0.3805     &      \\
Ethnicity: Non-white                    & 0.3499 & 0.3144 & 264.0  & 1.113    & 0.2667     &      \\
\bottomrule
\end{tabularx}
\label{tab:fixation_counts_results}
\end{table}

\begin{figure}[ht]
    \centering
    \includegraphics[width=0.9\linewidth]{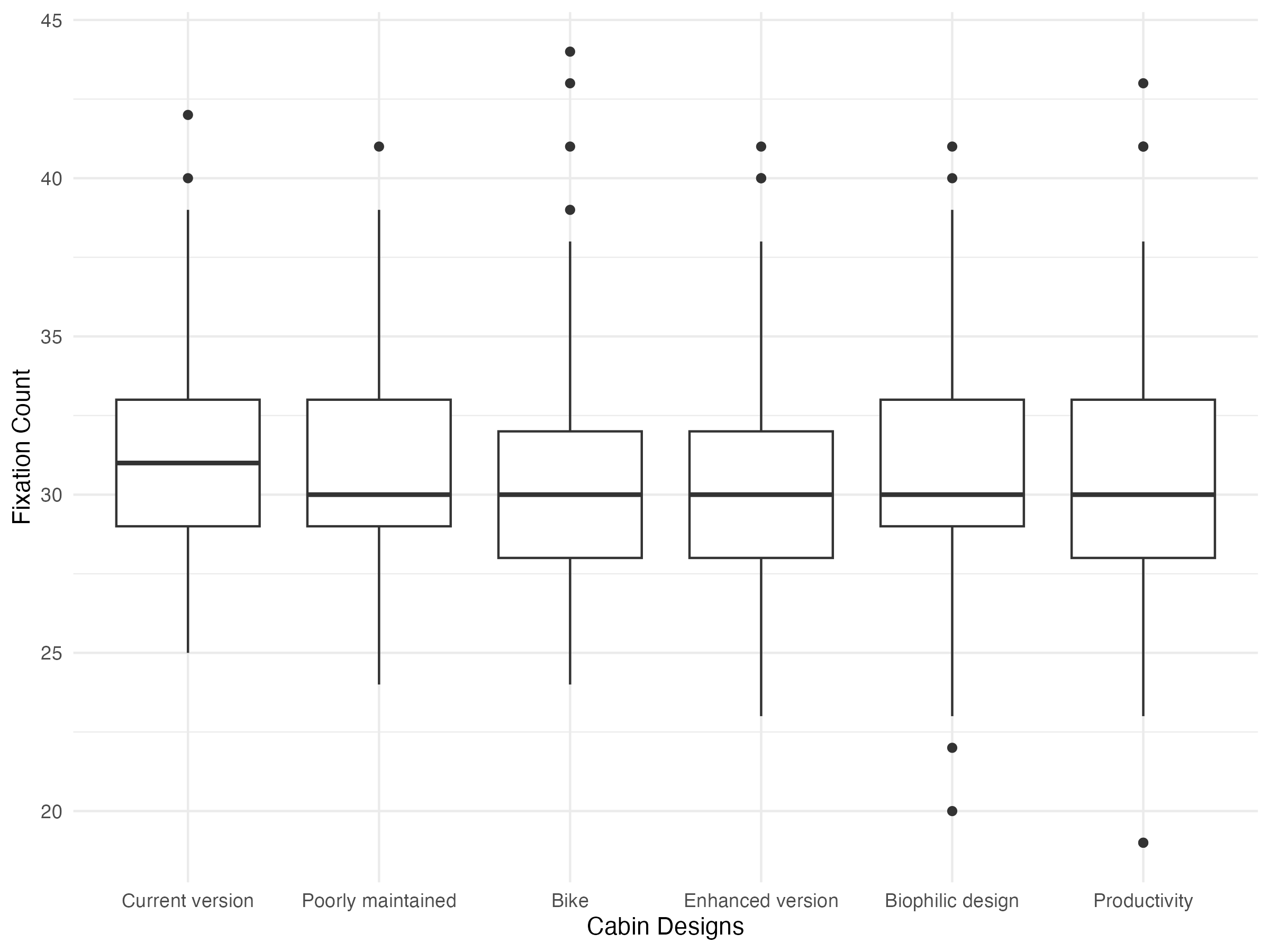} 
    \caption{Boxplot showing the distribution of fixation counts across different images.}
    \label{fig:fixation_counts_Boxplot}
\end{figure}

\subsubsection{Time to First Fixation}

On the other hand, a comparison of Time to First Fixation among the different cabin designs was carried out using a linear mixed effects model. Unlike fixation counts, the results indicate that all tested designs elicited statistically significant reductions in TFF, when compared to the baseline intercept (\textit{Current version}). Among the designs, the \textit{Biophilic design} demonstrated the largest negative effect (-1443.00 ms, \(p < 2 \times 10^{-16}\)). Similarly, the \textit{Poorly maintained current version}, \textit{Bike}, \textit{Productivity}, and \textit{Enhanced version} designs also exhibited large and statistically significant reductions in TFF (\(p < 2 \times 10^{-16}\) for all). These findings suggest that participants oriented their gaze more rapidly to each of these alternative designs than to the baseline, drawing their attention.

\begin{table}[ht]
\centering
\caption{Comparison of Time to first fixation across different cabin designs using a linear mixed effect model}
\label{tab:TFF}
\begin{tabular}{@{}lcccccc@{}}
\toprule
\textbf{Fixed Effects}                  & \textbf{Estimate} & \textbf{Std. Error} & \textbf{df}     & \textbf{t-value} & \textbf{p-value}       \\ \midrule
(Intercept)                             & 1476.31           & 49.89               & 1613.72         & 29.59            & \textless{}2e-16 ***   \\
Poorly maintained current version                & -1420.86          & 69.15               & 1361.43         & -20.55           & \textless{}2e-16 ***   \\
Bike                      & -1441.62          & 69.15               & 1361.43         & -20.85           & \textless{}2e-16 ***   \\
Productivity                    & -1419.08          & 69.15               & 1361.43         & -20.52           & \textless{}2e-16 ***   \\
Biophilic design              & -1443.00          & 69.15               & 1361.43         & -20.87           & \textless{}2e-16 ***   \\
Enhanced version                       & -1425.63          & 69.15               & 1361.43         & -20.62           & \textless{}2e-16 ***   \\ \bottomrule
\end{tabular}

\begin{flushleft}
\textbf{Significance Codes:} 0 ‘***’ 0.001 ‘**’ 0.01 ‘*’ 0.05 ‘.’ 0.1 ‘ ’ 1
\end{flushleft}
\end{table}

\begin{figure}[ht]
    \centering
    \includegraphics[width=0.9\linewidth]{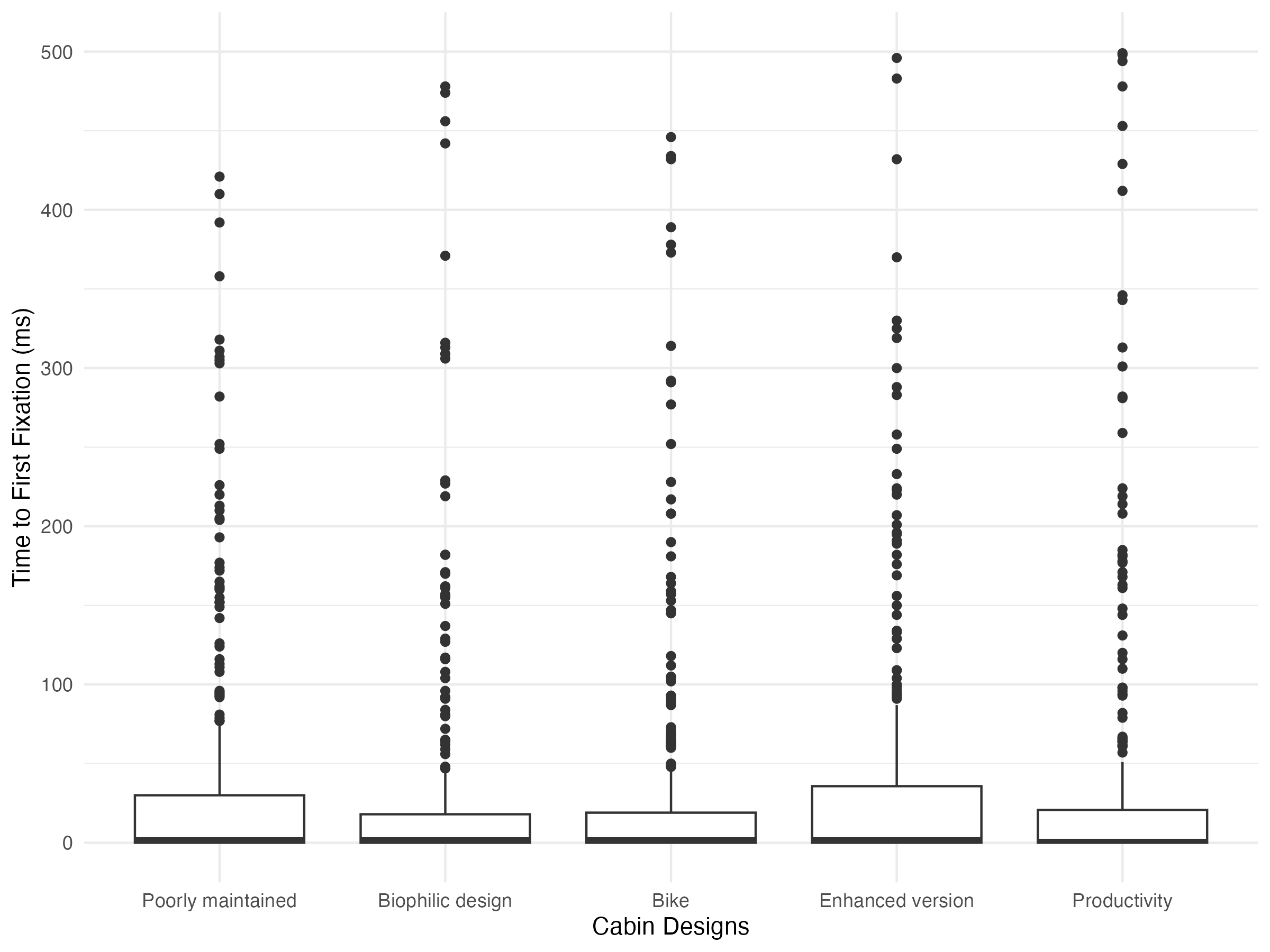} 
    \caption{Time to first fixation across different images. The boxplot shows the distribution of time to first fixation for each image.}
    \label{fig:TFF_Boxplot}
\end{figure}

\subsubsection{First Fixation Duration}
A comparison of first fixation durations among the different cabin designs was carried out using a linear mixed effects model. The results indicate that none of the tested designs elicited statistically significant differences in FFD, when compared to the baseline intercept (\textit{Current version}). Among the designs, the \textit{Biophilic design} demonstrated the largest positive effect (16.414 ms) and approached significance (\(p = 0.0544\)), though it did not meet the conventional threshold for statistical significance (\(p < 0.05\)). Similarly, other designs, including \textit{Poorly maintained current version}, \textit{Bike}, \textit{Productivity}, and \textit{Enhanced version}, exhibited minor positive or negative effects, but none achieved statistical significance (\(p > 0.05\)). These findings suggest that participants did not substantially differentiate among the designs in terms of their initial fixation durations, although the possibility of differences cannot be ruled out, potentially due to limited sample size or variability in the data.


\begin{table}[h!]
\centering
\caption{Comparison of First Fixation Duration across different cabin designs using a linear mixed effect model}
\label{tab:FFD}
\begin{tabular}{@{}lccccc@{}}
\toprule
\textbf{Variable} & \textbf{Estimate} & \textbf{Std. Error} & \textbf{df} & \textbf{t-value} & \textbf{p-value} \\ 
\midrule
(Intercept) & 276.604 & 6.972 & 1310.493 & 39.676 & \textless{}2e-16*** \\
Poorly maintained current version & -1.158 & 8.525 & 1349.569 & -0.136 & 0.8920 \\
Bike & 10.179 & 8.525 & 1349.569 & 1.194 & 0.2326 \\
Productivity & 11.004 & 8.525 & 1349.569 & 1.291 & 0.1970 \\
Biophilic design & 16.414 & 8.525 & 1349.569 & 1.925 & 0.0544. \\
Enhanced version & 11.330 & 8.525 & 1349.569 & 1.329 & 0.1841 \\
\bottomrule
\end{tabular}
\end{table}


\begin{figure}[ht]
    \centering
    \includegraphics[width=0.9\linewidth]{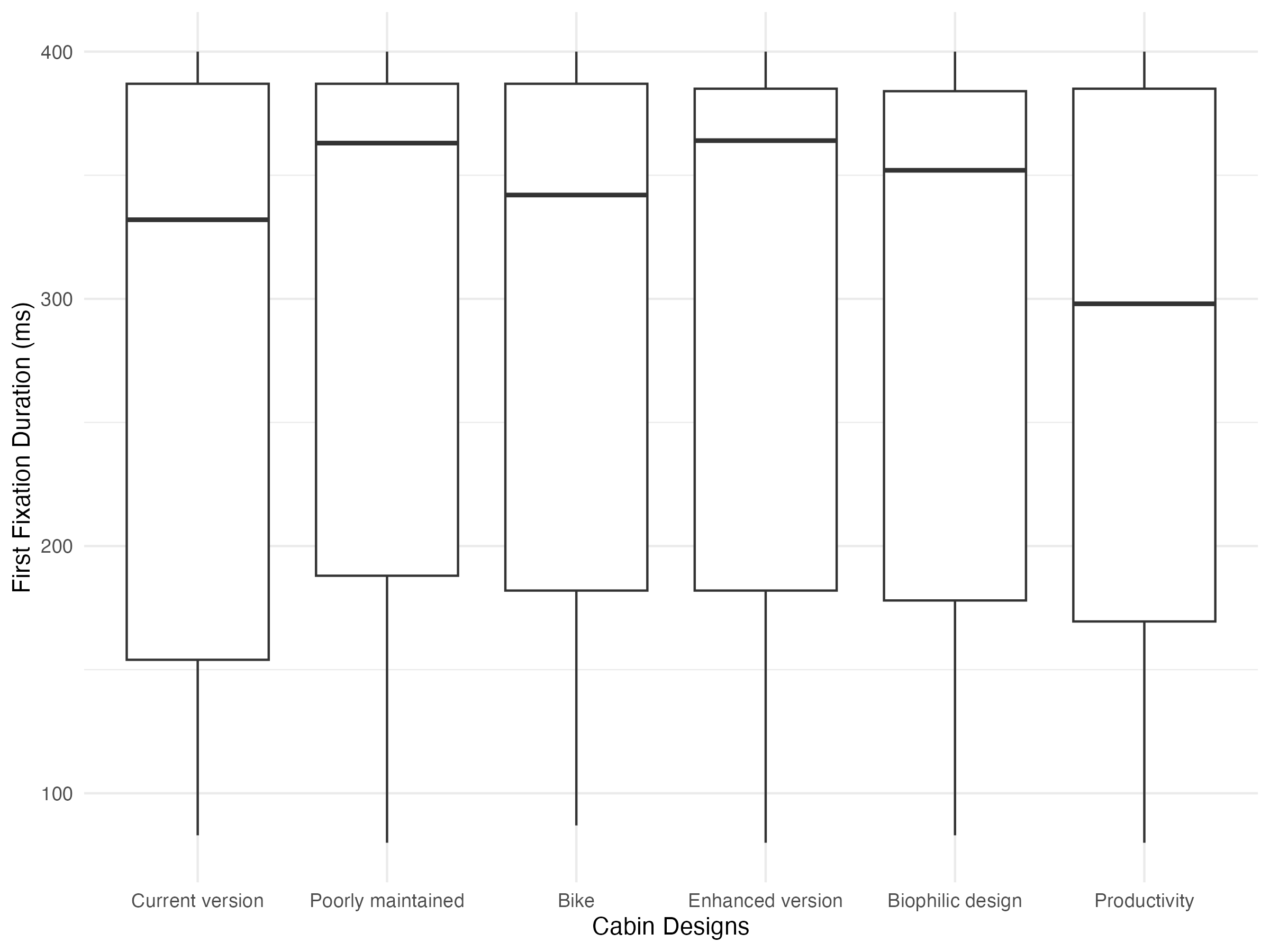} 
    \caption{First Fixation Duration across different images. The boxplot shows the distribution of first fixation duration for each image.}
    \label{fig:FFD_Boxplot}
\end{figure}

\subsubsection{Stationary Gaze Entropy}



An analysis of stationary gaze entropy (SGE) among the different cabin designs was carried out using a linear mixed effects model. The results, presented in Table \ref{tab:sge_result}, indicate that several tested designs elicited significant deviations in SGE relative to the baseline intercept (\textit{Current version}). Among the designs, the \textit{Bike} configuration exhibited the largest negative effect (-0.191, \(p < 2 \times 10^{-16}\)), reflecting a substantial reduction in gaze dispersion. Similarly, the \textit{Biophilic design} (-0.124, \(p < 2 \times 10^{-16}\)), \textit{Productivity} (-0.067, \(p < 2 \times 10^{-16}\)), and \textit{Enhanced version} (-0.168, \(p < 2 \times 10^{-16}\)) conditions all demonstrated significant decreases in SGE, suggesting more focused and predictable gaze behavior. In contrast, the \textit{Poorly maintained current version} elicited a slight but significant increase in SGE (0.012, \(p = 0.00705\)), indicating a more dispersed gaze pattern compared to the baseline. These findings suggest that certain cabin designs may promote more concentrated visual engagement, while others lead to more scattered gaze patterns.

\begin{table}[ht]
\centering
\caption{Linear Mixed Model Results for the Stationary Gaze Entropy in all conditions }
\begin{tabularx}{\textwidth}{@{}Xcccccc@{}}
\toprule
\textbf{Variable} & \textbf{Estimate} & \textbf{Std. Error} & \textbf{df} & \textbf{t value} & \textbf{Pr(>|t|)} \\
\midrule
Intercept                              & 3.973e+00 & 2.700e-02 & 2.705e+02 & 147.139 & <2e-16*** \\
Poorly maintained current version  & 7.845e-03 & 4.575e-03 & 4.880e+04 & 1.715   & 0.0864. \\
Bike       & -1.944e-01 & 4.589e-03 & 4.880e+04 & -42.374 & <2e-16*** \\
Productivity      & -6.728e-02 & 4.577e-03 & 4.880e+04 & -14.700 & <2e-16*** \\
Biophilic design  & -1.251e-01 & 4.585e-03 & 4.880e+04 & -27.276 & <2e-16*** \\
Enhanced version          & -1.699e-01 & 4.590e-03 & 4.880e+04 & -37.012 & <2e-16*** \\
Gender: Non-male                       & 2.503e-02 & 3.637e-02 & 2.640e+02 & 0.688   & 0.4919 \\
Ethnicity: Non-white                   & 1.480e-02 & 3.981e-02 & 2.640e+02 & 0.372   & 0.7103 \\
\bottomrule
\end{tabularx}
\label{tab:sge_result}
\end{table}

\begin{figure}[ht]
    \centering
    \includegraphics[width=0.9\linewidth]{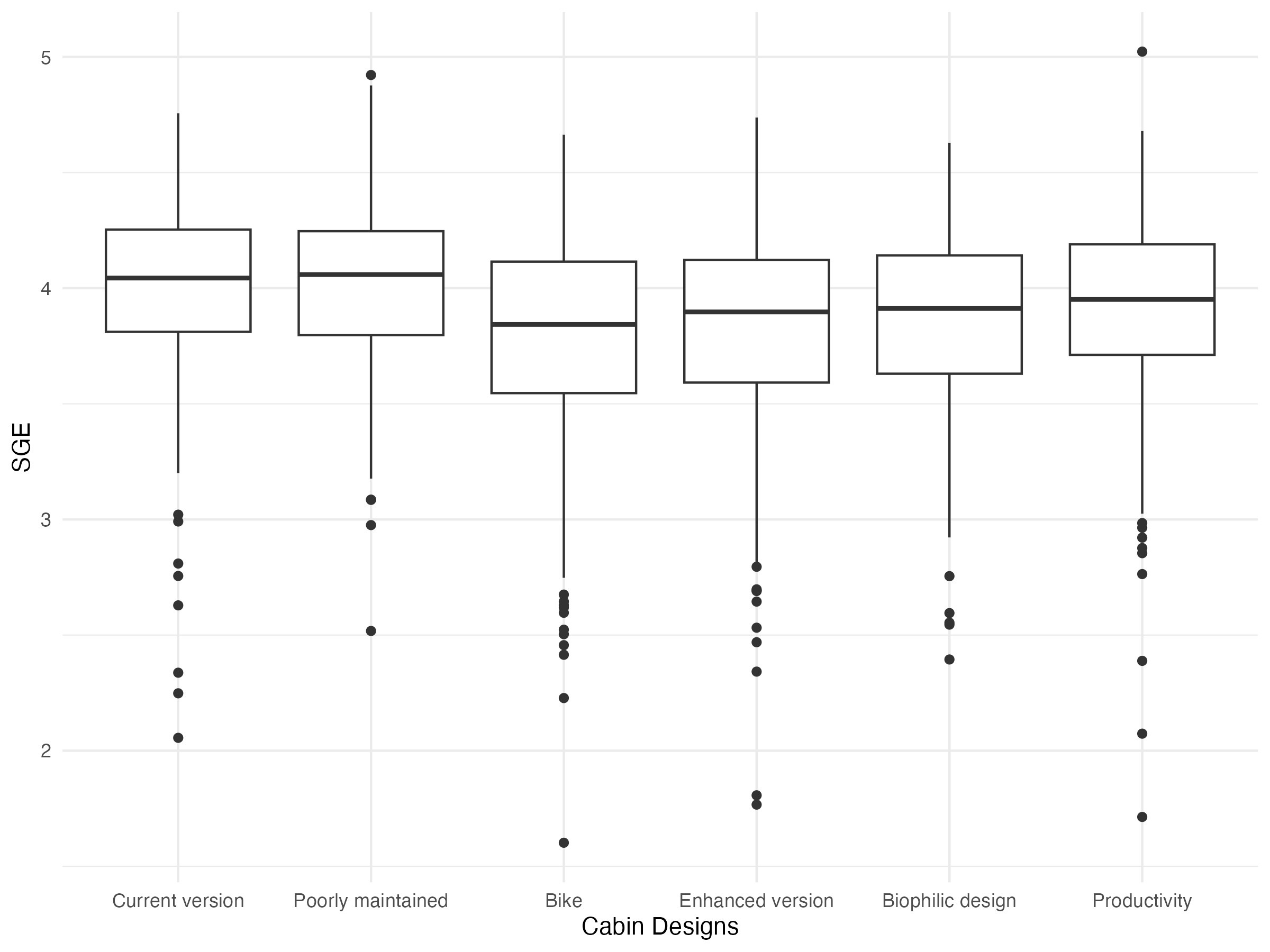} 
    \caption{Stationary Gaze Entropy across different images. The boxplot shows the distribution of SGE for each image.}
    \label{fig:SGE_boxplot}
\end{figure}


\subsubsection{Gaze Transitionary Entropy}

A comparison of Gaze Transition Entropy among the different cabin designs was carried out using a linear mixed effects model. As shown in Table \ref{tab:lmm_gte_item_filename}, the intercept (4.781) represents the baseline GTE for participants in the reference condition (\textit{Current version}), indicating a relatively high degree of randomness in gaze transitions under this condition. Among the designs, the \textit{Enhanced version} demonstrated the largest negative effect (-0.056, \(p < 2 \times 10^{-16}\)), reflecting a substantial decrease in gaze transition randomness. Similarly, the \textit{Bike} (-0.053, \(p < 2 \times 10^{-16}\)), \textit{Biophilic design} (-0.025, \(p < 2 \times 10^{-16}\)), and \textit{Productivity} (-0.008, \(p = 0.00483\)) conditions also exhibited significant reductions in GTE, suggesting more structured and predictable gaze transitions. In contrast, the \textit{Poorly maintained current version} elicited a slight but significant increase in GTE (0.017, \(p = 5.61 \times 10^{-10}\)), indicating a more scattered gaze pattern compared to the baseline. These findings imply that designs featuring more intentional, streamlined visual elements may promote focused and predictable gaze behavior, while environments with higher visual complexity or clutter may encourage more dispersed and less directed gaze transitions.

\begin{table}[ht]
\centering
\caption{Linear Mixed Effect Model Results for the Gaze Transitionary Entropy in all conditions}
\begin{tabularx}{\textwidth}{@{}Xcccccc@{}}
\toprule
\textbf{Variable} & \textbf{Estimate} & \textbf{Std. Error} & \textbf{df} & \textbf{t value} & \textbf{Pr(>|t|)} \\
\midrule
Intercept                              & 4.767e+00 & 1.636e-02 & 2.706e+02 & 291.428 & <2e-16*** \\
Poorly maintained current version  & 1.411e-02 & 2.810e-03 & 4.880e+04 & 5.020   & 5.17e-07*** \\
Bike        & -5.533e-02 & 2.819e-03 & 4.880e+04 & -19.631 & <2e-16*** \\
Productivity      & -7.461e-03 & 2.812e-03 & 4.880e+04 & -2.654  & 0.00797** \\
Biophilic design   & -2.601e-02 & 2.816e-03 & 4.880e+04 & -9.236  & <2e-16*** \\
Enhanced version          & -5.781e-02 & 2.820e-03 & 4.880e+04 & -20.502 & <2e-16*** \\
Gender: Non-male                       & 2.635e-02 & 2.203e-02 & 2.640e+02 & 1.196   & 0.23269 \\
Ethnicity: Non-white                   & 1.589e-02 & 2.411e-02 & 2.640e+02 & 0.659   & 0.51036 \\
\bottomrule
\end{tabularx}
\label{tab:lmm_gte_item_filename}
\end{table}


\begin{figure}[ht]
    \centering
    \includegraphics[width=0.9\linewidth]{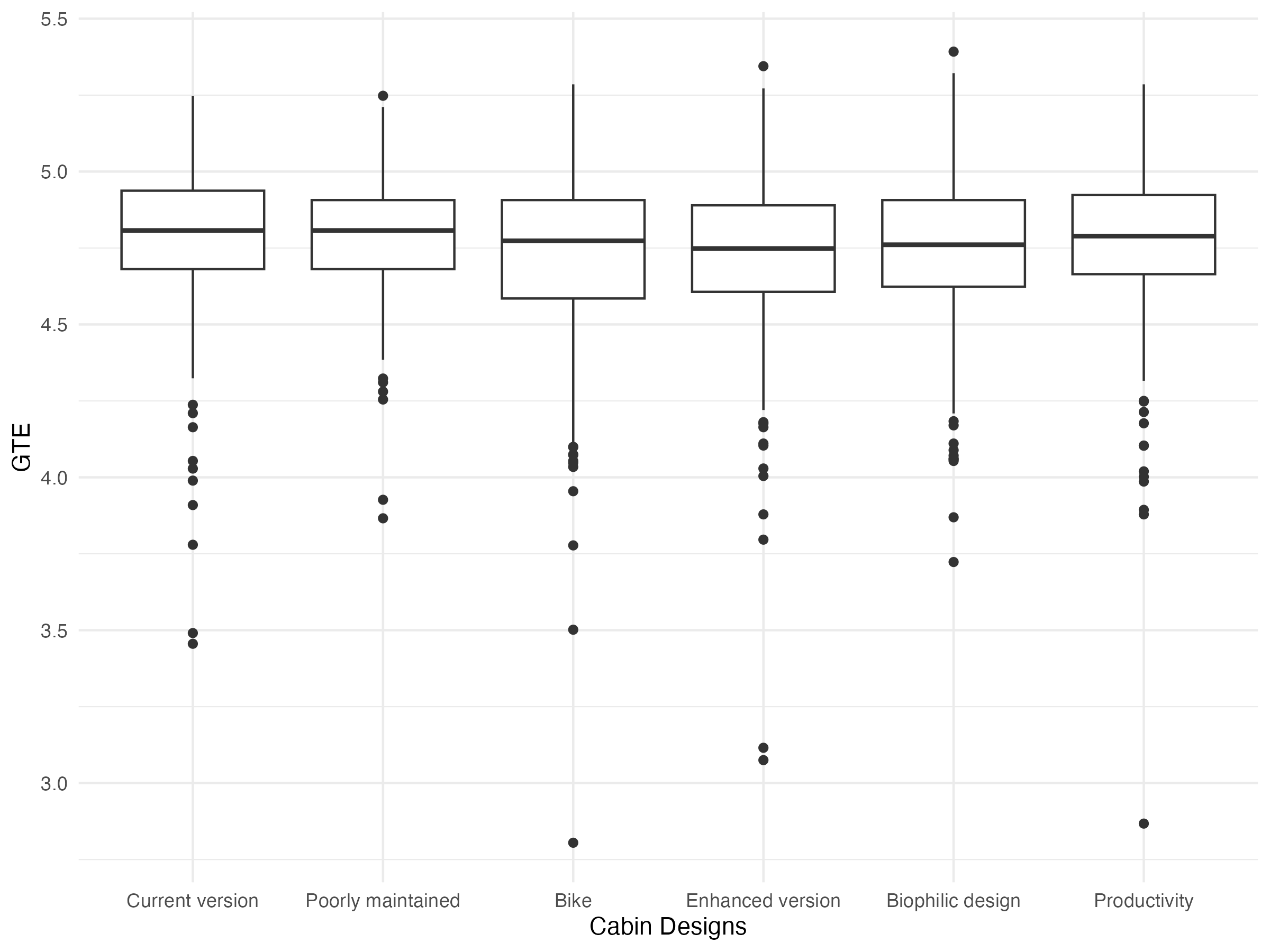} 
    \caption{Gaze Transitionary Entropy across different images. The boxplot shows the distribution of GTE for each image.}
    \label{fig:GTE_boxplot}
\end{figure}






\subsection{Emotional State Demographic Variables}\label{sec:emo}

We investigated the influence of participants' emotional states before viewing the cabin designs as a proxy for their emotional state prior entering public transportation, specifically stress, valence, and arousal on their visual engagement, as measured by Time to First Fixation, First Fixation Duration and Entropy measures across the cabin designs. The regression analyses did not reveal any statistically significant relationships between these emotional state variables and the eye-tracking metrics. These findings suggest that baseline emotional states may not substantially impact initial visual attention in the context of observing cabin environments. For a comprehensive overview of the regression results, please refer to Appendix~\ref{sec:appendix_emo}.


\subsection{Transportation Demographic Variables}\label{sec:demo}
In the following section we will provide the results of analysis for the relationship between demographic variables (i.e., Frequency of Public Transportation Use, Duration of Public Transport Use, and Purpose of Trip) and eye tracking metrics (i.e., First Fixation Duration, Time To First Fixation, Stationary Gaze Entropy, and Gaze Transitionary Entropy). To better structure this section, we analyze each cabin separately. For brevity, among the eye tracking metrics of interest, we will only present the ones that showed significant results with respect to the transportation demographics of interest.

\subsubsection{Frequency of Public Transport Use and First Fixation Duration}
For frequency of public transport use, we grouped responses into two broader categories: \textit{Most of the Time}, which combined the original responses of \textit{Always}, and \textit{Most of the Time}, and \textit{About Half of the Time and Less}, which included the responses of \textit{About Half of the Time}, \textit{Sometimes}, and \textit{Never}. This section focuses exclusively on the First Fixation Duration metric from the eye-tracking data, as other measures did not yield significant results in our modeling efforts. Overall, the regression analyses revealed that public transport use frequency did not significantly affect FFD in the \textit{Current Version}, \textit{Enhanced}, and \textit{Bike-centered} cabin designs. However, in the \textit{Poorly Maintained} and \textit{Biophilic} designs, participants who used public transport about half of the time or less exhibited significantly shorter FFDs compared to more frequent users. Additionally, in the \textit{Productivity} cabin design, ethnicity emerged as a significant predictor of FFD, indicating that demographic factors may influence visual engagement in certain environments. The detail of the analysis is provided below.

\paragraph{Current Version}

\noindent A linear regression model was used to examine the effect of public transport use frequency on the first fixation duration in the \textit{Current version} condition, while also controlling for gender and ethnicity. As shown in Table~\ref{tab:FFD_current_frequency}, none of the included variables produced statistically significant effects on FFD. Specifically, participants who used public transport ``About half of the time and less'' displayed no significant difference in FFD compared to those who used it ``Most of the time'' (Estimate = -2.720, \(p = 0.885\)). Similarly, identifying as Non-male (Estimate = 1.228, \(p = 0.932\)) or Non-white (Estimate = -13.139, \(p = 0.406\)) was not associated with any significant variation in FFD relative to the baseline categories. These findings should be interpreted with caution, as the lack of significant effects may stem from methodological constraints such as limited sample size or insufficient variability.

\begin{table}[ht]
\centering
\caption{Regression Results for the Effect of Frequency of Public Transport Use on the First Fixation Duration in ``Current version''}
\begin{tabularx}{\textwidth}{@{}Xcccc@{}}
\toprule
\textbf{Variable} & \textbf{Estimate} & \textbf{Std. Error} & \textbf{t value} & \textbf{Pr(>|t|)} \\
\midrule
Intercept (Baseline: Male, White, Most of the time) & 279.433 & 19.460 & 14.359 & <0.0001*** \\
frequency: About half of the time and less & -2.720 & 18.847 & -0.144 & 0.885 \\
Gender: Non-male & 1.228 & 14.439 & 0.085 & 0.932 \\
Ethnicity: Non-white & -13.139 & 15.792 & -0.832 & 0.406 \\
\bottomrule
\end{tabularx}
\label{tab:FFD_current_frequency}
\end{table}

\begin{figure}[ht]
    \centering
    \includegraphics[width=0.9\linewidth]{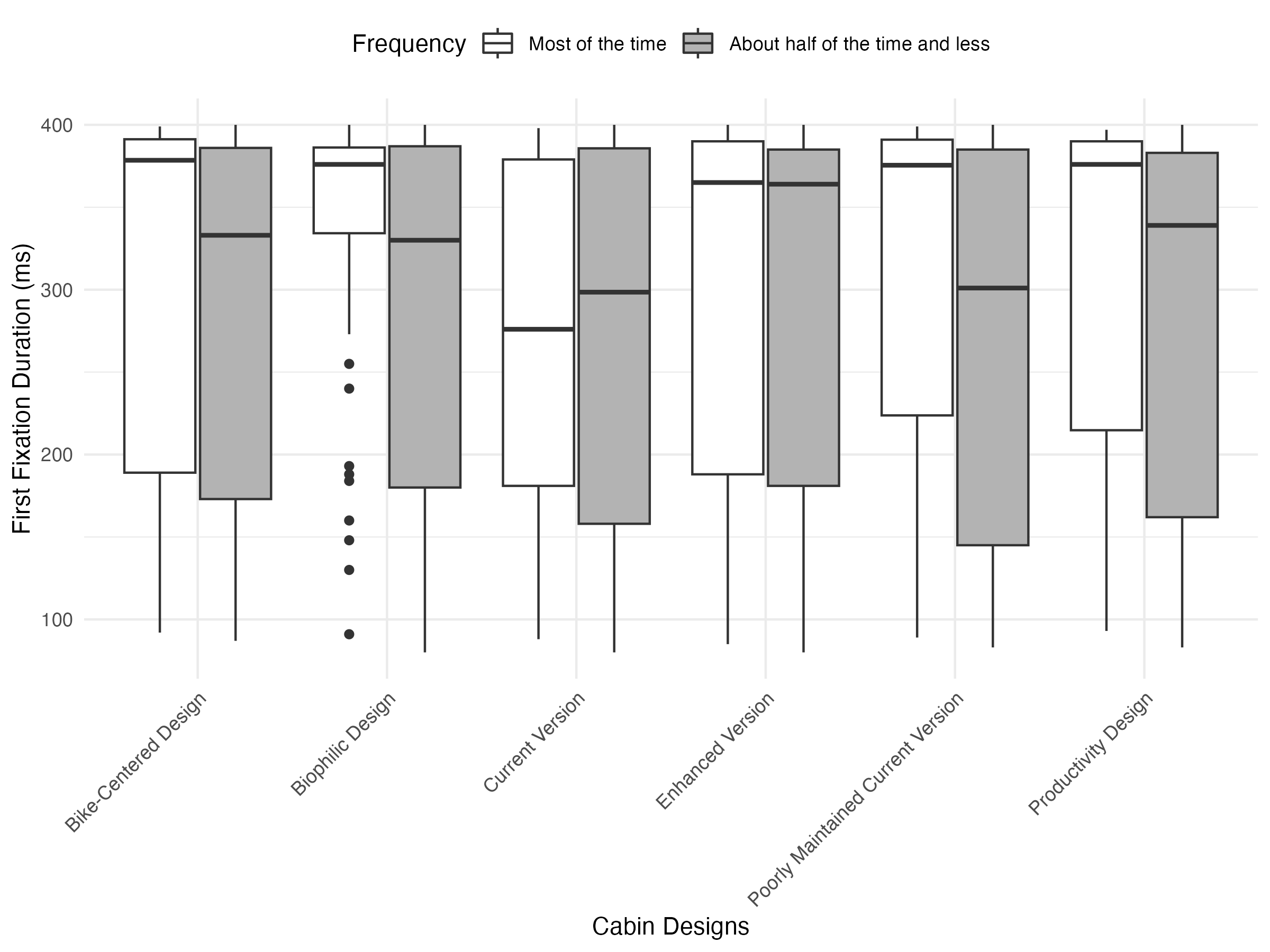} 
    \caption{Distribution of First Fixation Duration By Frequency of Public Transport Use}
    \label{fig:frequency_FFD}
\end{figure}

\paragraph{Poorly Maintained} 
As shown in Table~\ref{tab:baseline_trash_frequency}, the regression analysis reveals a statistically significant association between public transport use frequency and the First Fixation Duration in the \textit{Poorly maintained current version} condition. Participants who reported using public transport about half of the time or less exhibited a 46.96~ms decrease in FFD \((p = 0.014)\), suggesting that they directed their gaze more rapidly than those who use public transport more frequently. Additionally, identifying as Non-white was associated with a 34.36~ms reduction in FFD \((p = 0.032)\).

\begin{table}[ht]
\centering
\caption{Regression Results for the effect of frequency of public transport use on the first fixation duration in ``Poorly maintained current version''}
\begin{tabularx}{\textwidth}{@{}Xcccc@{}}
\toprule
\textbf{Variable} & \textbf{Estimate} & \textbf{Std. Error} & \textbf{t value} & \textbf{Pr(>|t|)} \\
\midrule
Intercept (Baseline: Male, White, Most of the time) & 323.033 & 19.561 & 16.514 & <0.0001*** \\
frequency: About half of the time and less & -46.960 & 18.983 & -2.474 & 0.0140* \\
Gender: Non-male & -1.587 & 14.484 & -0.110 & 0.9129 \\
Ethnicity: Non-white & -34.363 & 15.930 & -2.157 & 0.0319* \\
\bottomrule
\end{tabularx}
\label{tab:baseline_trash_frequency}
\end{table}

\paragraph{Enhanced version} 

Table~\ref{tab:FFD_enhanced_frequency} presents the regression results for the \textit{Enhanced version} cabin design. Similar to the \textit{Bike-centered} condition, public transport use frequency did not have a statistically significant effect on First Fixation Duration. Although participants who used public transport about half of the time or less showed a slightly shorter FFD (-11.672~ms), this difference was not significant \((p = 0.520)\).

\begin{table}[ht]
\centering
\caption{Regression Results for the Effect of Frequency of Public Transport Use on the First Fixation Duration in ``Enhanced version'' cabin design}
\begin{tabularx}{\textwidth}{@{}Xcccc@{}}
\toprule
\textbf{Variable} & \textbf{Estimate} & \textbf{Std. Error} & \textbf{t value} & \textbf{Pr(>|t|)} \\
\midrule
Intercept (Baseline: Male, White, Most of the time) & 309.945 & 18.675 & 16.597 & <0.0001*** \\
frequency: About half of the time and less & -11.672 & 18.124 & -0.644 & 0.520 \\
Gender: Non-male & -9.664 & 13.828 & -0.699 & 0.485 \\
Ethnicity: Non-white & -28.727 & 15.209 & -1.889 & 0.060 . \\
\bottomrule
\end{tabularx}
\label{tab:FFD_enhanced_frequency}
\end{table}

\paragraph{Biophilic Design}

As shown in Table~\ref{tab:wellbeing_plants_frequency}, the regression analysis for the \textit{Biophilic} cabin design reveals a statistically significant effect of public transport use frequency on First Fixation Duration. Participants who used public transport about half of the time or less exhibited a 53.39~ms reduction in FFD \((p = 0.00218)\), suggesting that they oriented their gaze more rapidly than those who use public transport more frequently. This finding may indicate that less frequent users are more sensitive to the biophilic elements or find them more novel, resulting in quicker initial engagement.


\begin{table}[ht]
\centering
\caption{Regression Results for the effect of frequency of public transport use on the first fixation duration in ''Biophilic'' cabin design}
\begin{tabularx}{\textwidth}{@{}Xcccc@{}}
\toprule
\textbf{Variable} & \textbf{Estimate} & \textbf{Std. Error} & \textbf{t value} & \textbf{Pr(>|t|)} \\
\midrule
Intercept (Baseline: Male, White, Most of the time) & 339.56 & 17.77 & 19.104 & <0.0001*** \\
frequency: About half of the time and less & -53.39 & 17.25 & -3.095 & 0.00218** \\
Gender: Non-male & 3.40 & 13.16 & 0.258 & 0.79634 \\
Ethnicity: Non-white & -16.59 & 14.47 & -1.146 & 0.25273 \\
\bottomrule
\end{tabularx}
\label{tab:wellbeing_plants_frequency}
\end{table}

\paragraph{Bike} 

As shown in Table~\ref{tab:FDD_bike_frequency}, the regression analysis for the \textit{Bike-centered} cabin design did not reveal a statistically significant effect of public transport use frequency on First Fixation Duration. Although the estimate for participants who use public transport about half of the time or less was negative (-20.69~ms), this difference did not approach conventional levels of significance \((p = 0.243)\).

\begin{table}[ht]
\centering
\caption{Regression Results for the Effect of Frequency of Public Transport Use on the First Fixation Duration in ``Bike-centered'' cabin design}
\begin{tabularx}{\textwidth}{@{}Xcccc@{}}
\toprule
\textbf{Variable} & \textbf{Estimate} & \textbf{Std. Error} & \textbf{t value} & \textbf{Pr(>|t|)} \\
\midrule
Intercept (Baseline: Male, White, Most of the time) & 319.87 & 18.21 & 17.563 & <0.0001*** \\
frequency: About half of the time and less & -20.69 & 17.67 & -1.171 & 0.243 \\
Gender: Non-male & -20.83 & 13.49 & -1.544 & 0.124 \\
Ethnicity: Non-white & -20.85 & 14.83 & -1.406 & 0.161 \\
\bottomrule
\end{tabularx}
\label{tab:FDD_bike_frequency}
\end{table}

\paragraph{Productivity}

\noindent As shown in Table~\ref{tab:FFD_productivity_frequency}, the regression analysis for the \textit{Productivity} cabin design did not yield a statistically significant effect of public transport use frequency on First Fixation Duration. Participants who reported using public transport about half of the time or less had an FFD that was shorter by 26.77~ms, but this difference was not significant \((p = 0.139)\). In contrast, identifying as Non-white was associated with a 29.92~ms reduction in FFD, which reached statistical significance \((p = 0.049)\). These results highlight that demographic factors may play a role in visual engagement.

\begin{table}[ht]
\centering
\caption{Regression Results for the Effect of Frequency of Public Transport Use on the First Fixation Duration in ``Productivity'' cabin design}
\begin{tabularx}{\textwidth}{@{}Xcccc@{}}
\toprule
\textbf{Variable} & \textbf{Estimate} & \textbf{Std. Error} & \textbf{t value} & \textbf{Pr(>|t|)} \\
\midrule
Intercept (Baseline: Male, White, Most of the time) & 311.27 & 18.57 & 16.761 & <0.0001*** \\
frequency: About half of the time and less & -26.77 & 18.02 & -1.485 & 0.1387 \\
Gender: Non-male & 11.99 & 13.75 & 0.872 & 0.3839 \\
Ethnicity: Non-white & -29.92 & 15.12 & -1.978 & 0.0489* \\
\bottomrule
\end{tabularx}
\label{tab:FFD_productivity_frequency}
\end{table}

\subsubsection{Purpose of the Trip and Time to First Fixation}
For purpose of the trip variable, we focus exclusively on the time to first fixation metric from the eye-tracking data, as other measures did not yield significant results in our modeling efforts for any of the cabin designs when considering the effect of demographic variables. Overall, the regression analyses indicated that the purpose of the trip did not significantly affect TFF in the \textit{Current Version}, \textit{Poorly Maintained}, \textit{Biophilic}, \textit{Bike-centered}, and \textit{Productivity} cabin designs. However, in the \textit{Enhanced} cabin design, both commuting to work and leisure travel purposes were significantly associated with shorter TFFs compared to other usage patterns. Additionally, in the \textit{Biophilic} cabin design, ethnicity emerged as a significant predictor of TFF, suggesting that demographic factors may influence visual engagement in specific environments. The detail of the analysis is provided below.

\paragraph{Current Version}

As shown in Table~\ref{tab:tff_purpose_baseline}, the linear regression analysis examining the effect of public transport use purpose on Time to First Fixation in the \textit{Current version} environment did not reveal any statistically significant relationships. Neither commuting to work (Estimate = 52.50~ms, \(p = 0.9483\)) nor leisure travel (Estimate = -37.17~ms, \(p = 0.9626\)) significantly influenced TFF.

\begin{table}[ht]
\centering
\caption{Regression Results for the effect of purpose of public transportation use on the time to first fixation in the ``Current version'' }
\begin{tabularx}{\textwidth}{@{}Xcccc@{}}
\toprule
\textbf{Variable} & \textbf{Estimate} & \textbf{Std. Error} & \textbf{t value} & \textbf{Pr(>|t|)} \\
\midrule
Intercept (Baseline: Shopping, Male, White) & 1559.10 & 787.19 & 1.981 & 0.0502 . \\
Commute to Work                                    & 52.50   & 808.13 & 0.065 & 0.9483   \\
Leisure                                           & -37.17  & 790.49 & -0.047 & 0.9626   \\
Gender: Non-male                                  & -520.30 & 352.75 & -1.475 & 0.1432   \\
Ethnicity: Non-white                              & 461.15  & 405.69 & 1.137 & 0.2582   \\
\bottomrule
\end{tabularx}
\label{tab:tff_purpose_baseline}
\end{table}

\paragraph{Poorly Maintained}

As shown in Table~\ref{tab:tff_purpose_trash}, the linear regression analysis for the \textit{Poorly maintained current version} environment did not reveal any statistically significant effects of public transport use purpose on Time to First Fixation. Neither commuting to work (Estimate = -2.692~ms, \(p = 0.937\)) nor leisure use (Estimate = -26.442~ms, \(p = 0.421\)) was associated with a statistically significant change in TFF.

\begin{figure}[ht]
    \centering
    \includegraphics[width=0.9\linewidth]{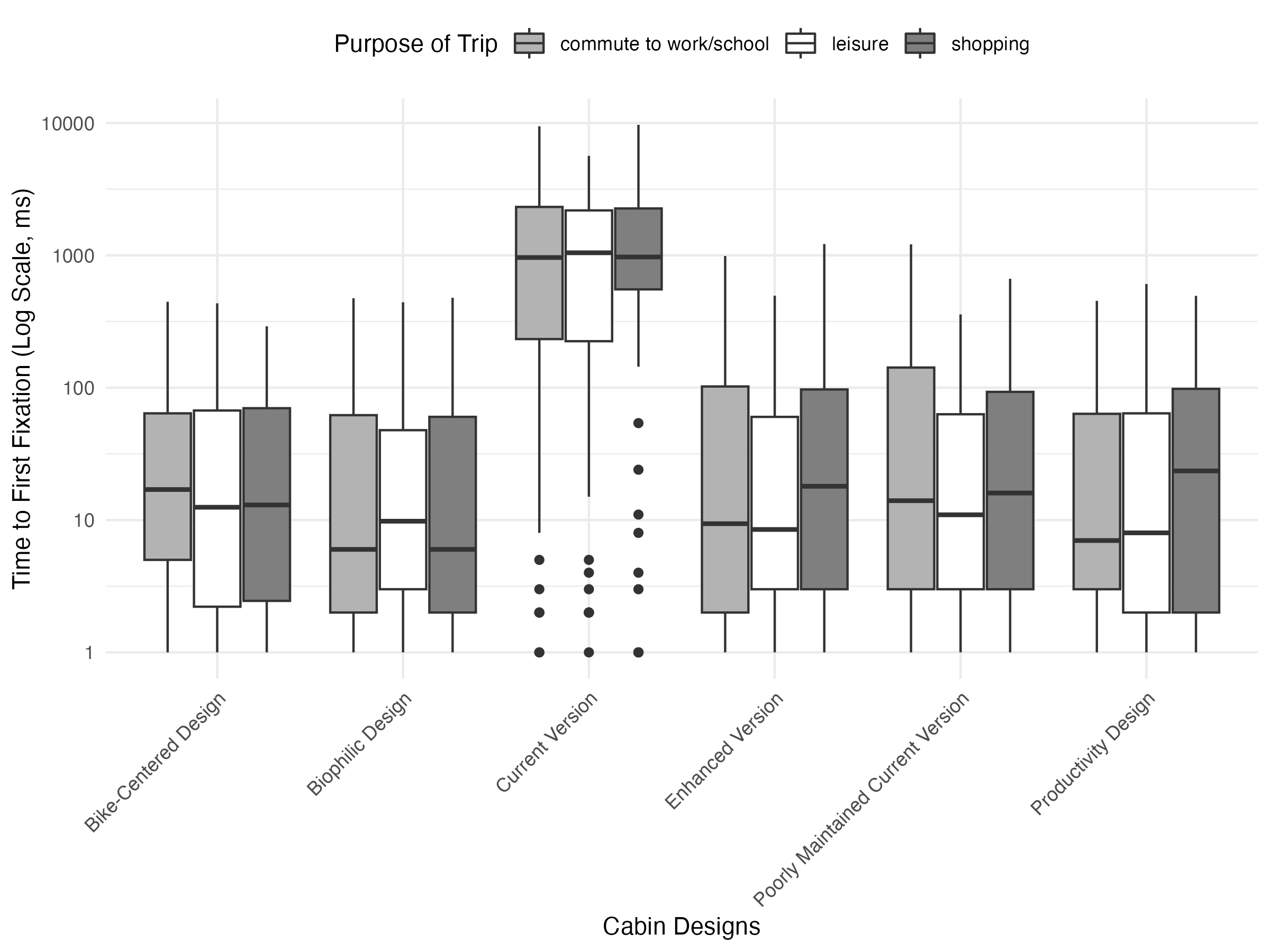} 
    \caption{Distribution of Time to First Fixation By Purpose of Public Transport Use}
    \label{fig:purpose_TFF}
\end{figure}

\begin{table}[ht]
\centering
\caption{Regression Results for the effect of purpose of public transportation use on the time to first fixation in the ``Poorly maintained current version''}
\begin{tabularx}{\textwidth}{@{}Xcccc@{}}
\toprule
\textbf{Variable} & \textbf{Estimate} & \textbf{Std. Error} & \textbf{t value} & \textbf{Pr(>|t|)} \\
\midrule
Intercept (Baseline: Shopping, Male, White)   & 52.855 & 32.917 & 1.606 & 0.111 \\
Commute to Work    
& -2.692 & 33.871 & -0.079 & 0.937 \\
Leisure                                   & -26.442 & 32.756 & -0.807 & 0.421 \\
Gender: Non-male                          & 12.544 & 16.160 & 0.776 & 0.439 \\
Ethnicity: Non-white                      & 15.719 & 18.812 & 0.836 & 0.405 \\
\bottomrule
\end{tabularx}
\label{tab:tff_purpose_trash}
\end{table}

\paragraph{Enhanced Version}

\noindent As shown in Table~\ref{tab:tff_purpose_enhanced}, the regression analysis for the \textit{Enhanced} cabin design revealed a significant association between the purpose of public transport use and Time to First Fixation. Participants who commute to work (-139.20~ms, \(p = 0.0203\)) or use public transport for leisure (-140.77~ms, \(p = 0.0154\)) exhibited significantly shorter TFFs compared to those with other usage patterns. These results suggest that travel purpose may influence how quickly individuals orient their gaze in a more visually streamlined environment.

\begin{table}[ht]
\centering
\caption{Regression Results for the effect of purpose of public transportation use on the time to first fixation in the ``Enhanced'' cabin design}
\begin{tabularx}{\textwidth}{@{}Xcccc@{}}
\toprule
\textbf{Variable} & \textbf{Estimate} & \textbf{Std. Error} & \textbf{t value} & \textbf{Pr(>|t|)} \\
\midrule
Intercept (Baseline: Shopping, Male, White) & 138.43 & 57.49 & 2.408 & 0.0176* \\
Commute to Work                                     & -139.20 & 59.16 & -2.353 & 0.0203* \\
Leisure                                            & -140.77 & 57.21 & -2.461 & 0.0154* \\
Gender: Non-male                                   & 55.88 & 28.22 & 1.980 & 0.0501. \\
Ethnicity: Non-white                               & 85.56 & 32.86 & 2.604 & 0.0104* \\
\bottomrule
\end{tabularx}
\label{tab:tff_purpose_enhanced}
\end{table}

\paragraph{Biophilic Design}

\noindent As shown in Table~\ref{tab:tff_purpose_plants}, the regression analysis for the \textit{Biophilic} cabin design did not yield any statistically significant effects of public transport use purpose on Time to First Fixation. Neither commuting to work (Estimate = 13.808~ms, \(p = 0.604\)) nor leisure travel (Estimate = 7.022~ms, \(p = 0.785\)) showed a meaningful impact on TFF. Although identifying as Non-white was associated with a significant increase in TFF \((p < 0.001)\), the absence of significant effects for travel purpose should be interpreted with caution, considering potential limitations such as sample size or variability in the data.

\begin{table}[ht]
\centering
\caption{Regression Results for the effect of purpose of public transportation use on the time to first fixation in the ``Biophilic'' cabin design}
\begin{tabularx}{\textwidth}{@{}Xcccc@{}}
\toprule
\textbf{Variable} & \textbf{Estimate} & \textbf{Std. Error} & \textbf{t value} & \textbf{Pr(>|t|)} \\
\midrule
Intercept (Baseline: Shopping, Male, White) & 2.046 & 25.777 & 0.079 & 0.937 \\
Commute to Work                                     & 13.808 & 26.523 & 0.521 & 0.604 \\
Leisure                                            & 7.022 & 25.650 & 0.274 & 0.785 \\
Gender: Non-male                                   & 22.377 & 12.654 & 1.768 & 0.080 . \\
Ethnicity: Non-white                               & 56.873 & 14.731 & 3.861 & <0.001*** \\
\bottomrule
\end{tabularx}
\label{tab:tff_purpose_plants}
\end{table}

\paragraph{Bike}

\noindent As shown in Table~\ref{tab:tff_purpose_bike}, the regression analysis for the \textit{Bike-centered} cabin design did not reveal any statistically significant relationship between the purpose of public transport use and Time to First Fixation. Neither commuting to work (Estimate = 17.234~ms, \(p = 0.607\)) nor leisure travel (Estimate = 28.888~ms, \(p = 0.373\)) significantly influenced TFF.

\begin{table}[ht]
\centering
\caption{Regression Results for the effect of purpose of public transportation use on the time to first fixation in the ``Bike-centered'' cabin design}
\begin{tabularx}{\textwidth}{@{}Xcccc@{}}
\toprule
\textbf{Variable} & \textbf{Estimate} & \textbf{Std. Error} & \textbf{t value} & \textbf{Pr(>|t|)} \\
\midrule
Intercept (Baseline: Shopping, Male, White) & 13.239 & 32.427 & 0.408 & 0.684 \\
Commute to Work                                     & 17.234 & 33.367 & 0.516 & 0.607 \\
Leisure                                            & 28.888 & 32.268 & 0.895 & 0.373 \\
Gender: Non-male                                   & -0.212 & 15.919 & -0.013 & 0.989 \\
Ethnicity: Non-white                               & -1.926 & 18.532 & -0.104 & 0.917 \\
\bottomrule
\end{tabularx}
\label{tab:tff_purpose_bike}
\end{table}

\paragraph{Productivity}

\noindent As shown in Table~\ref{tab:tff_purpose_productivity}, the regression analysis for the \textit{Productivity} cabin design did not reveal any statistically significant effects of public transport use purpose on Time to First Fixation. Neither commuting to work (Estimate = -27.801~ms, \(p = 0.4687\)) nor leisure travel (Estimate = -24.051~ms, \(p = 0.5168\)) significantly influenced TFF. Although identifying as Non-white was associated with a significant increase in TFF \((p < 0.0002)\), the overall results suggest that the purpose of public transport use did not meaningfully shape participants’ initial visual engagement with this cabin.

\begin{table}[ht]
\centering
\caption{Regression Results for the effect of purpose of public transportation use on the time to first fixation in the ``Productivity'' cabin design}
\begin{tabularx}{\textwidth}{@{}Xcccc@{}}
\toprule
\textbf{Variable} & \textbf{Estimate} & \textbf{Std. Error} & \textbf{t value} & \textbf{Pr(>|t|)} \\
\midrule
Intercept (Baseline: Shopping, Male, White) & 41.132 & 37.165 & 1.107 & 0.2707 \\
Commute to Work                                     & -27.801 & 38.242 & -0.727 & 0.4687 \\
Leisure                                            & -24.051 & 36.983 & -0.650 & 0.5168 \\
Gender: Non-male                                   & 9.726 & 18.245 & 0.533 & 0.5950 \\
Ethnicity: Non-white                               & 83.159 & 21.240 & 3.915 & <0.0002*** \\
\bottomrule
\end{tabularx}
\label{tab:tff_purpose_productivity}
\end{table}

\subsubsection{Duration of Public Transport Use}

This analysis examines the impact of public transport use duration on First Fixation Duration across various cabin designs. Across all cabin designs—including \textit{Current Version}, \textit{Poorly Maintained}, \textit{Enhanced}, \textit{Biophilic}, \textit{Bike-centered}, and \textit{Productivity}—the regression analyses consistently indicated no statistically significant relationships between public transport use duration and FFD. The duration categories (e.g., 15--30 min, 30--45 min, 45--60 min, ``Don’t know,'' and more than one hour) did not produce notable deviations in FFD compared to the reference condition (all \(p > 0.05\)). A detailed analysis of the results is provided as an appendix in Section \ref{sec:app1}.

While the majority of results were non-significant, certain demographic variables exhibited significant effects in specific cabin designs:
\begin{itemize}
    \item \textbf{Poorly Maintained Current Version:} Identifying as Non-white was associated with a statistically significant reduction in FFD (\(p = 0.00593\)).
    \item \textbf{Enhanced Version:} Identifying as Non-white remained a significant predictor, associated with a 41.66 ms reduction in FFD (\(p = 0.0147\)).
\end{itemize}

In order to access a consolidated view of the regression results across all cabin designs, refer to Table~\ref{tab:consolidated_fdd_duration}.

\begin{table}[ht]
\centering
\caption{Consolidated Regression Results for the Effect of Public Transportation Use Duration on First Fixation Duration (FFD) Across Cabin Designs}
\begin{tabularx}{\textwidth}{@{}lccc@{}}
\toprule
\textbf{Cabin Design} & \textbf{Duration Effect} & \textbf{Gender Effect} & \textbf{Ethnicity Effect} \\ \midrule
Current Version        & None                     & None                     & None                          \\
Poorly Maintained      & None                     & None                     & Non-white (\(p = 0.00593\))   \\
Enhanced               & None                     & None                     & Non-white (\(p = 0.0147\))    \\
Biophilic              & None                     & None                     & None                          \\
Bike-centered          & None                     & None                     & None (\(p = 0.0822\))          \\
Productivity           & None                     & None                     & None                          \\
\bottomrule
\end{tabularx}
\label{tab:consolidated_fdd_duration}
\end{table}

\begin{figure}[ht]
    \centering
    \includegraphics[width=0.9\linewidth]{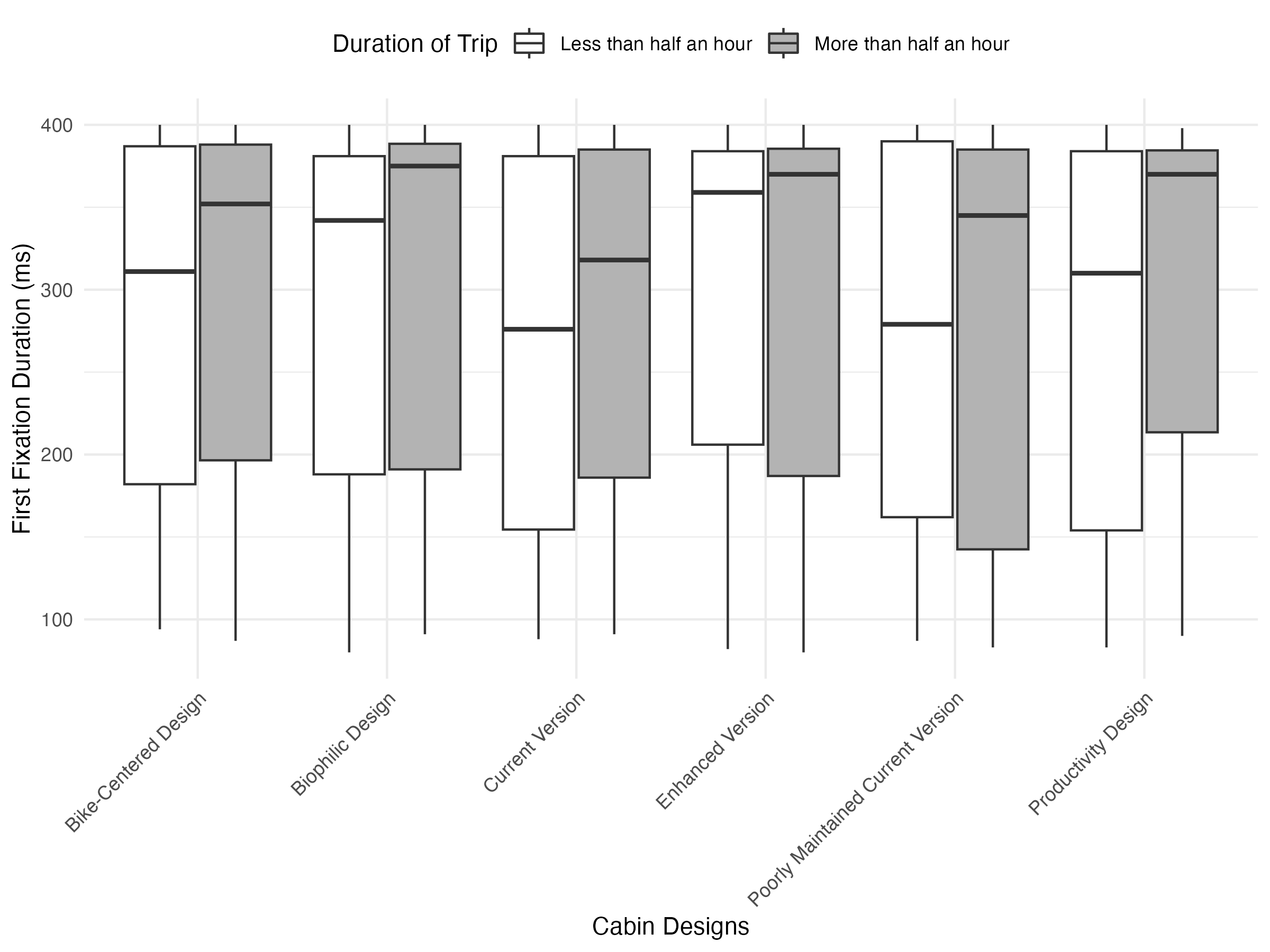} 
    \caption{Distribution of First Fixation Duration By Duration of Public Transport Use}
    \label{fig:duration_FFD}
\end{figure}

\section{Discussion}

This study advances our understanding of how passengers visually engage with public transportation cabin environments by integrating eye-tracking metrics with public transportation usage factors. While earlier research primarily relied on self-reported surveys \cite{hakiminejad2024public, imam2014measuring, del2013determining}, our approach offers a more direct and nuanced perspective on how passengers visually navigate these spaces. By analyzing gaze behavior across groups defined by various usage patterns and demographic factors, we uncover subtle relationships that may be overlooked by self-reported measures alone. Our findings demonstrate that while certain cabin designs and demographic variables can influence the manner and speed with which participants direct their gaze, these effects are neither universal nor consistent across all measured dimensions. In the following discussion, we highlight key observations for each eye-tracking metric, consider their potential implications for improving cabin design, and examine how participant characteristics—such as frequency, purpose, and duration of public transportation use—shape visual engagement patterns.

The comparisons of fixation counts among the various cabin designs revealed no statistically significant differences relative to the baseline \textit{Current Version} (Table~\ref{tab:fixation_counts_results}). Although the \textit{Enhanced Version} approached significance, the lack of a definitive effect may suggest that participants did not fundamentally vary their distribution of fixations based solely on the cabin design. One interpretation is that fixation count alone may be insufficiently sensitive to detect subtle differences in how participants visually parse a complex environment. Similar results were observed in previous studies with eye tracking data, where fixation counts did not adequately capture nuanced attentional shifts in intricate visual settings \cite{land2009looking}.

In contrast, the Time to First Fixation results presented a more pronounced pattern (Table~\ref{tab:TFF}) and revealed a different story. All tested designs, including \textit{Biophilic Design}, \textit{Poorly Maintained Current Version}, \textit{Bike}, \textit{Productivity}, and \textit{Enhanced Version}, demonstrated significantly shorter TFFs compared to the \textit{Current Version}. This pattern suggests that participants oriented their gaze more rapidly to features within these alternative configurations. Inline with prior literature, one possible explanation is that these designs contain more salient or attention-grabbing elements, guiding participants’ gaze more efficiently \cite{koch1987shifts,itti2001computational}. Prior studies show that environments incorporating biophilic design elements improve participants' perceptions of comfort and reduce stress levels \cite{altaf2023time,douglas2022physical,yin2019effects,yin2020effects,gaekwad2022meta}. These results were also replicated for cabin designs by increasing comfort, decreasing stress, and negative emotions \cite{hakiminejad2024public}. The eye-tracking results support these findings by indicating that such designs contain highly salient features that facilitate intuitive visual engagement. For example, previous findings revealed that participants rated biophilic cabins as significantly more calming and inviting, which aligns with the shorter TFF observed here, suggesting that participants quickly focused on familiar and visually appealing elements like greenery and natural textures. Our findings demonstrate how both conscious impressions and unconscious gaze behaviors highlight the importance of design choices that foster positive user experiences.

For First Fixation Duration, no design exhibited a statistically significant difference compared to the \textit{Current Version}, though the \textit{Biophilic Design} approached significance (Table~\ref{tab:FFD}). FFD reflects the initial temporal investment of attention on the first element viewed, and the lack of significant differences implies that once participants found their initial fixation point, they did not spend substantially more or less time there across different designs. This outcome further underscores the complexity of interpreting eye-tracking metrics in isolation. The rapid orientation suggested by TFF findings does not necessarily translate into deeper attention as measured by FFD.

The entropy-based metrics, SGE and GTE, provide a more holistic measure of visual exploration patterns. The results indicated that certain cabin designs led to more structured and predictable gaze distributions (lower entropy), while others encouraged more scattered gaze patterns (higher entropy) (Tables~\ref{tab:sge_result} and \ref{tab:lmm_gte_item_filename}). Specifically, the \textit{Bike}, \textit{Biophilic Design}, \textit{Productivity}, and \textit{Enhanced Version} designs all resulted in significantly lower SGE and GTE, suggesting that participants visually parsed these environments in a more systematic and coherent manner, which aligns with prior findings demonstrating that structured environments facilitate more efficient visual processing \cite{henderson1999high}.  These patterns may reflect a design’s visual clarity, thematic coherence, or presence of salient features that help guide participants’ gaze. Conversely, the \textit{Poorly Maintained Current Version} condition resulted in increased entropy measures, potentially indicating visual clutter or elements that distract participants from forming a stable viewing pattern. These findings imply that designs featuring more intentional, streamlined visual elements promote focused and predictable gaze behavior, enhancing the efficiency with which passengers navigate the space. Lower entropy in gaze patterns can be indicative of environments that facilitate easier information processing and reduce cognitive load, contributing to a more comfortable and user-friendly experience \cite{shiferaw2019review}. These findings collectively suggest that intentional and streamlined visual elements in cabin designs not only attract quicker attention but also facilitate more organized visual processing, enhancing the efficiency with which passengers navigate the space.

Beyond cabin design, we explored how demographic variables—specifically frequency, purpose, and duration of public transportation use—relate to eye-tracking metrics. These factors offer insights into how individual differences and travel behaviors influence visual engagement with cabin environments. For frequency of public transport use, most cabin designs did not show a significant relationship with FFD, except for a few noteworthy cases. In the \textit{Poorly Maintained Current Version} and \textit{Biophilic Design} conditions, participants who used public transport about half of the time or less exhibited significantly shorter FFDs compared to more frequent users (Tables~\ref{tab:baseline_trash_frequency} and \ref{tab:wellbeing_plants_frequency}). These findings point out that less frequent users might find certain elements more novel or striking, prompting quicker initial fixations. Alternatively, familiarity with public transport settings could influence the expectation of what to observe first, leading frequent users to scan more systematically rather than locking onto salient features immediately.

Regarding the purpose of the trip, the regression analyses indicated that the purpose of the trip did not significantly affect TFF in most cabin designs, including \textit{Current Version}, \textit{Poorly Maintained}, \textit{Biophilic}, \textit{Bike-centered}, and \textit{Productivity} (Table~\ref{tab:consolidated_fdd_duration}). However, in the \textit{Enhanced Version} cabin design, both commuting to work and leisure travel purposes were significantly associated with shorter TFFs compared to other usage patterns (Table~\ref{tab:tff_purpose_enhanced}). This suggests that in more visually streamlined environments, the purpose of travel can influence how quickly individuals orient their gaze, potentially reflecting different attentional strategies based on travel motivations.

The analysis of public transport use duration revealed that, overall, duration did not significantly influence FFD across most cabin designs (Table~\ref{tab:consolidated_fdd_duration}). However, ethnicity emerged as a significant predictor of FFD in several cabin environments. Specifically, in the \textit{Productivity} cabin design, participants identifying as Non-white exhibited a significant reduction in FFD (Table~\ref{tab:FFD_productivity_frequency}). Additionally, in both the \textit{Enhanced Version} and \textit{Poorly Maintained Current Version} cabin designs, Non-white participants demonstrated shorter FFDs (Tables~\ref{tab:tff_purpose_enhanced} and \ref{tab:FFD_productivity_frequency}). These demographic differences suggest that ethnicity may influence visual attention patterns within certain cabin environments, potentially reflecting diverse visual processing strategies or varying levels of familiarity with the design elements. This finding underscores the importance of considering demographic diversity in cabin design to ensure that visual elements cater to a broad spectrum of users.

While frequency and purpose of public transport use generally did not impact FFD and TFF across all designs, significant interactions in specific cabin environments highlight the nuanced ways in which demographic and behavioral factors intersect with design elements to shape visual engagement. These interactions suggest that passenger characteristics can modulate the effectiveness of design features in directing visual attention, emphasizing the need for inclusive and adaptable cabin designs.

\section{Implications for Cabin Design}

These findings point to several implications for public transport cabin design that can enhance both user experience and operational efficiency. First, maintaining cleanliness and reducing clutter is essential for minimizing visual distractions. Environments that resemble the \textit{Poorly Maintained Current Version} produce scattered gaze patterns and heightened mental effort as passengers attempt to navigate disordered spaces. By keeping interiors tidy, organized, and free from unnecessary signage or decorations, designers can help passengers quickly locate important features, fostering a sense of comfort and trust that may encourage continued ridership.

In addition to cleanliness, integrating natural elements and employing structured layouts can guide passenger attention more intuitively. Cabins inspired by the \textit{Biophilic Design} or \textit{Enhanced Version} concepts, for instance, were associated with more predictable gaze patterns and faster orientation. Incorporating greenery, uniform color schemes, and clearly delineated zones can help passengers efficiently process visual information, improving their perception of the environment’s aesthetics and functionality. This streamlined visual experience not only enhances user satisfaction but may also reinforce perceptions of safety and well-being \cite{hakiminejad2024public}.

Demographic considerations further inform these design strategies. Since less frequent or first-time public transport users may struggle to orient themselves, the aforementioned design can enhance their experience. By ensuring that diverse passenger groups—from daily commuters to occasional leisure travelers—can quickly and confidently interpret the environment, transportation systems can become more inclusive and appealing. Additionally, recognizing that ethnicity can influence visual engagement patterns, designers should consider culturally diverse perspectives to create universally navigable and comfortable spaces.

Crucially, these implications need not remain static. Continuous refinement through feedback—collected via passenger surveys, focus groups, and pilot tests—can guide incremental improvements. By complementing traditional qualitative insights with objective, real-time measures obtained through eye-tracking, our methodology establishes a feedback loop that connects genuine user perceptions with tangible design adaptations. This approach allows designers to test new interventions, observe how users genuinely respond in real-world scenarios, and then refine the space accordingly. In doing so, public transport cabin environments can evolve steadily, becoming not only cleaner, more organized, and more intuitive, but also finely attuned to passengers’ evolving expectations and needs. Over time, this iterative process holds the potential to make public transportation more accessible, enjoyable, and appealing to a wide range of users, ultimately advancing broader sustainability and mobility objectives.

\section{Limitations and Future Directions}

While our findings provide valuable insights into how passengers visually engage with different public transportation cabin designs, several limitations should be noted. First, the study was conducted using \textit{static images} rather than immersive or real-world conditions. Although eye-tracking data from these images can highlight participants’ initial visual orientation, it may not fully capture the dynamic nature of real transit environments—where movement, crowding, and ambient factors (e.g., noise, temperature) can influence attention and perception. Future research could employ virtual reality (VR) headsets or augmented reality (AR) simulations to create more ecologically valid scenarios, enabling participants to navigate and interact with virtual cabin spaces in real time.

Second, the within-subjects design, while controlling for individual differences in baseline visual behavior, could introduce \textit{familiarity effects}. Although the image presentation was randomized, repeated exposure to multiple cabin designs may have led participants to become more familiar with the task and change their exploration patterns (e.g., scanning more quickly or focusing on different elements) by the final designs. Employing between-subjects designs or counterbalancing the order of stimuli across subgroups may help mitigate these potential carryover effects.

Third, the demographic breakdown of the sample—while diverse—may not fully represent broader transit-riding populations, especially regarding age, health, and socio-economic diversity. Factors such as visual acuity, disability, cultural background, or experience with other transit systems can mediate user responses. Additionally, participants were recruited via an online platform (Prolific), which may skew toward individuals who are more technologically savvy or have reliable internet access. Future investigations could integrate on-site or lab-based data collection to sample passengers with different technology access or specific ridership profiles (e.g., older adults, individuals with visual or cognitive impairments) for more inclusive insights.

Fourth, while the remote eye-tracking approach enabled broad participation and flexible data collection, it inherently introduces variability—such as differences in monitor size, lighting conditions, or webcam quality—that may affect tracking accuracy. Although the RealEye platform includes calibration procedures and data-quality metrics, minor inaccuracies could still influence the results (e.g., drifting fixations in edge regions of the screen). Employing standardized hardware in controlled laboratory settings or using portable eye-tracking glasses for field studies can improve data precision and validate these remote findings.

Fifth, the study focused on relatively short (10-second) exposures to each cabin image. Although this duration is useful for capturing quick, initial impressions and “bottom-up” visual salience, longer viewing times might reveal deeper exploration strategies—such as searching for signage, noticing comfort features, or assessing cleanliness more thoroughly. Future work could vary exposure lengths or allow participants to freely explore 360° cabin panoramas to better simulate how actual passengers might scan a space over the course of a journey.

Finally, while the entropy-based metrics (Stationary Gaze Entropy and Gaze Transition Entropy) offer rich insights into the structure of visual exploration, they do not directly capture emotional responses or interpret whether a gaze pattern is “positive” or “negative.” Integrating psychophysiological measures (e.g., skin conductance, heart rate variability), observational data (e.g., dwell times near certain amenities in an actual transit car), or qualitative interviews could enrich our understanding of the interplay between visual engagement, emotional reactions, and ultimate behavior (e.g., willingness to ride). Longitudinal studies, following participants over repeated encounters with newly implemented cabin features, could also ascertain whether initial attention patterns predict real shifts in satisfaction, perceived safety, and ridership.

By addressing these methodological and contextual constraints, future research can build upon our findings to develop more comprehensive and ecologically valid evaluations of public transportation environments. Doing so will not only improve cabin design strategies but also help meet broader goals of increasing transit adoption, enhancing user well-being, and promoting more sustainable mobility.



\section{Conclusion}

Our study contributes to a nuanced understanding of how cabin design influences visual attention and exploration. While the \textit{Current version} did not strongly differ from other designs in fixation counts or FFD, significant patterns in TFF and entropy-based metrics emerged, painting a more intricate picture of engagement. Designs that introduce biophilic elements, enhanced organization, or thematic coherence appear to encourage more structured and focused gaze patterns, whereas cluttered or less optimized environments foster more dispersed viewing.

Demographic variables related to public transport use also demonstrated selective influences on gaze metrics, highlighting that not all individuals engage with these environments in the same manner. Such differences may reflect varying levels of familiarity, motivation, or susceptibility to environmental cues. Although many relationships did not reach statistical significance, the trends observed encourage further exploration into the interplay between user characteristics and environmental design.

Overall, these findings emphasize the importance of employing multiple eye-tracking metrics and considering participant heterogeneity when evaluating visual engagement in complex environments. By doing so, researchers and designers can gain richer insights into how to create or modify spaces that better guide attention, improve user experience, and ultimately influence perception and behavior.

\section{Acknowledgment}
This work received funding from Villanova University’s Falvey Memorial Library Scholarship Open Access Reserve (SOAR) Fund.


\bibliography{sn-bibliography}


\begin{appendices}

\section{Emotional State Demographic Variables}\label{sec:appendix_emo}

The results of the linear regression models examining the effects of participants' emotional states—specifically stress, valence, and arousal—on Time to First Fixation (TFF) and First Fixation Duration (FFD) across various cabin designs are presented in the following tables. While most analyses did not reveal statistically significant relationships, certain demographic factors interacted with emotional states to influence eye-tracking metrics. Detailed regression coefficients, standard errors, t-values, and p-values are provided for each emotional state variable and demographic factor to offer comprehensive insights into the data.

\subsubsection{Stress}

The regression analysis for stress on Time to First Fixation (TFF) within the \textit{Biophilic Design} cabin design indicated that ethnicity is a significant predictor, with Non-white participants exhibiting a shorter TFF compared to their white counterparts (\(p = 0.0119\)). However, baseline stress levels and gender did not show significant effects. Similarly, when assessing the impact of stress on First Fixation Duration (FFD) in the \textit{Poorly Maintained Current Version}, ethnicity remained a significant factor, with Non-white participants demonstrating a significant reduction in FFD (\(p = 0.00659\)). In the \textit{Enhanced} cabin design, ethnicity again significantly influenced FFD, while stress levels and gender did not show significant relationships.

\begin{table}[ht]
\centering
\caption{The Effect of Stress on the Time to First Fixation (TFF) in the ``Biophilic Design''}
\begin{tabularx}{\textwidth}{@{}lcccccc@{}}
\toprule
\textbf{Variable} & \textbf{Estimate} & \textbf{Std. Error} & \textbf{t value} & \textbf{Pr(>|t|)} & \textbf{Significance} \\
\midrule
(Intercept)         & 20.3284 & 11.4761 & 1.771 & 0.0781 & . \\
Pre\_Stress\_Survey & 0.3206  & 2.4572  & 0.130 & 0.8963 &   \\
Gender: Non-male    & 8.9632  & 12.5112 & 0.716 & 0.4746 &   \\
Ethnicity: Non-white & 33.2700 & 13.0996 & 2.540 & 0.0119 & * \\
\bottomrule
\end{tabularx}
\label{stress_tff_biophilic}
\end{table}

\begin{table}[ht]
\centering
\caption{The Effect of Stress on the First Fixation Duration (FFD) in the ``Poorly Maintained Current Version''}
\begin{tabularx}{\textwidth}{@{}lcccccc@{}}
\toprule
\textbf{Variable} & \textbf{Estimate} & \textbf{Std. Error} & \textbf{t value} & \textbf{Pr(>|t|)} & \textbf{Significance} \\
\midrule
(Intercept)           & 281.518 & 15.709 & 17.920 & $<$ 2e-16 & *** \\
Pre\_Stress\_Survey   & 1.151   & 3.364  & 0.342  & 0.73248   &     \\
Gender: Non-male      & 16.585  & 17.126 & 0.968  & 0.33408   &     \\
Ethnicity: Non-white  & -49.263 & 17.932 & -2.747 & 0.00659   & **  \\
\bottomrule
\end{tabularx}
\label{stress_ffd_trash}
\end{table}

\begin{table}[ht]
\centering
\caption{The Effect of Stress on the First Fixation Duration (FFD) in the ``Enhanced'' Cabin Design}
\begin{tabularx}{\textwidth}{@{}lcccccc@{}}
\toprule
\textbf{Variable} & \textbf{Estimate} & \textbf{Std. Error} & \textbf{t value} & \textbf{Pr(>|t|)} & \textbf{Significance} \\
\midrule
(Intercept)           & 310.317 & 14.734 & 21.062 & $<$ 2e-16 & *** \\
Pre\_Stress\_Survey   & -1.280  & 3.155  & -0.406 & 0.6855    &     \\
Gender: Non-male      & 1.840   & 16.062 & 0.115  & 0.9089    &     \\
Ethnicity: Non-white  & -41.168 & 16.818 & -2.448 & 0.0153    & *   \\
\bottomrule
\end{tabularx}
\label{stress_ffd_enhanced}
\end{table}

\subsubsection{Valence}

The analysis of valence on First Fixation Duration (FFD) in the \textit{Poorly Maintained Current Version} revealed that ethnicity significantly predicted FFD, with Non-white participants showing shorter fixation durations (\(p = 0.00596\)). Stress and gender did not significantly impact FFD in this design. In the \textit{Enhanced} cabin design, ethnicity again emerged as a significant predictor of FFD, while valence and gender were not significant. Additionally, in the \textit{Biophilic} cabin design, ethnicity significantly influenced Time to First Fixation (TFF), indicating that Non-white participants had shorter TFFs (\(p = 0.0129\)) compared to white participants.

\begin{table}[ht]
\centering
\caption{The Effect of Valence on the First Fixation Duration (FFD) in the ``Poorly Maintained Current Version''}
\begin{tabularx}{\textwidth}{@{}lcccccc@{}}
\toprule
\textbf{Variable} & \textbf{Estimate} & \textbf{Std. Error} & \textbf{t value} & \textbf{Pr(>|t|)} & \textbf{Significance} \\
\midrule
(Intercept)           & 309.184 & 29.442 & 10.501 & $<$ 2e-16 & *** \\
Pre\_Valence\_Survey  & -5.061  & 5.632  & -0.899 & 0.37002   &     \\
Gender: Non-male      & 17.158  & 16.957 & 1.012  & 0.31290   &     \\
Ethnicity: Non-white  & -49.817 & 17.911 & -2.781 & 0.00596   & **  \\
\bottomrule
\end{tabularx}
\label{valence_ffd_trash}
\end{table}

\begin{table}[ht]
\centering
\caption{The Effect of Valence on the First Fixation Duration (FFD) in the ``Enhanced'' Cabin Design}
\begin{tabularx}{\textwidth}{@{}lcccccc@{}}
\toprule
\textbf{Variable} & \textbf{Estimate} & \textbf{Std. Error} & \textbf{t value} & \textbf{Pr(>|t|)} & \textbf{Significance} \\
\midrule
(Intercept)           & 309.3412 & 27.6744 & 11.178 & $<$ 2e-16 & *** \\
Pre\_Valence\_Survey  & -0.6064  & 5.2939  & -0.115 & 0.9089    &     \\
Gender: Non-male      & 0.9874   & 15.9386 & 0.062  & 0.9507    &     \\
Ethnicity: Non-white  & -41.2765 & 16.8355 & -2.452 & 0.0151    & *   \\
\bottomrule
\end{tabularx}
\label{valence_ffd_enhanced}
\end{table}

\begin{table}[ht]
\centering
\caption{The Effect of Valence on the Time to First Fixation (TFF) in the ``Biophilic'' Cabin Design}
\begin{tabularx}{\textwidth}{@{}lcccccc@{}}
\toprule
\textbf{Variable} & \textbf{Estimate} & \textbf{Std. Error} & \textbf{t value} & \textbf{Pr(>|t|)} & \textbf{Significance} \\
\midrule
(Intercept)           & 39.103  & 21.502 & 1.819  & 0.0705 & .   \\
Pre\_Valence\_Survey  & -3.726  & 4.113  & -0.906 & 0.3661 &     \\
Gender: Non-male      & 9.042   & 12.384 & 0.730  & 0.4662 &     \\
Ethnicity: Non-white  & 32.846  & 13.081 & 2.511  & 0.0129 & *   \\
\bottomrule
\end{tabularx}
\label{valence_tff_plants}
\end{table}

\subsubsection{Arousal}

The regression analyses examining the impact of arousal on eye-tracking metrics revealed that ethnicity significantly influenced First Fixation Duration (FFD) in both the \textit{Poorly Maintained Current Version} (\(p = 0.00606\)) and the \textit{Enhanced} cabin designs (\(p = 0.0155\)), with Non-white participants showing reduced FFDs. However, baseline arousal levels and gender did not exhibit significant effects on FFD. Additionally, in the \textit{Biophilic} cabin design, ethnicity was a significant predictor of Time to First Fixation (TFF), indicating that Non-white participants had shorter TFFs (\(p = 0.0115\)). These findings suggest that while emotional states such as arousal do not directly impact visual metrics, demographic factors like ethnicity play a crucial role in moderating these relationships.

\begin{table}[ht]
\centering
\caption{The Effect of Arousal on the First Fixation Duration (FFD) in the ``Poorly Maintained Current Version''}
\begin{tabularx}{\textwidth}{@{}lcccccc@{}}
\toprule
\textbf{Variable} & \textbf{Estimate} & \textbf{Std. Error} & \textbf{t value} & \textbf{Pr(>|t|)} & \textbf{Significance} \\
\midrule
(Intercept)           & 303.438 & 24.330 & 12.472 & $<$ 2e-16 & *** \\
Pre\_Arousal\_Survey  & -4.719  & 5.426  & -0.870 & 0.38554   &     \\
Gender: Non-male      & 17.369  & 16.958 & 1.024  & 0.30703   &     \\
Ethnicity: Non-white  & -49.716 & 17.910 & -2.776 & 0.00606   & **  \\
\bottomrule
\end{tabularx}
\label{arousal_ffd_trash}
\end{table}

\begin{table}[ht]
\centering
\caption{The Effect of Arousal on the First Fixation Duration (FFD) in the ``Enhanced'' Cabin Design}
\begin{tabularx}{\textwidth}{@{}lcccccc@{}}
\toprule
\textbf{Variable} & \textbf{Estimate} & \textbf{Std. Error} & \textbf{t value} & \textbf{Pr(>|t|)} & \textbf{Significance} \\
\midrule
(Intercept)           & 302.635 & 22.864 & 13.236 & $<$ 2e-16 & *** \\
Pre\_Arousal\_Survey  & 0.975   & 5.099  & 0.191  & 0.8486    &     \\
Gender: Non-male      & 1.001   & 15.937 & 0.063  & 0.9500    &     \\
Ethnicity: Non-white  & -41.105 & 16.832 & -2.442 & 0.0155    & *   \\
\bottomrule
\end{tabularx}
\label{arousal_ffd_enhanced}
\end{table}

\begin{table}[ht]
\centering
\caption{The Effect of Arousal on the Time to First Fixation (TFF) in the ``Biophilic'' Cabin Design}
\begin{tabularx}{\textwidth}{@{}lcccccc@{}}
\toprule
\textbf{Variable} & \textbf{Estimate} & \textbf{Std. Error} & \textbf{t value} & \textbf{Pr(>|t|)} & \textbf{Significance} \\
\midrule
(Intercept)           & 14.835  & 17.796 & 0.834  & 0.4056 &     \\
Pre\_Arousal\_Survey  & 1.655   & 3.969  & 0.417  & 0.6772 &     \\
Gender: Non-male      & 9.159   & 12.404 & 0.738  & 0.4612 &     \\
Ethnicity: Non-white  & 33.451  & 13.100 & 2.553  & 0.0115 & *   \\
\bottomrule
\end{tabularx}
\label{arousal_tff_plants}
\end{table}

\section{Detailed Analysis of the Effect of Duration of Public Transport Use on FFD}\label{sec:app1}

\subsubsection{Duration of Public Transport Use}

For Duration of Public Transport Use variable, we focus exclusively on the time to first fixation metric from the eye-tracking data, as other measures did not yield significant results in our modeling efforts for any of the cabin designs when considering the effect of demographic variables.

\paragraph{Current Version}

As shown in Table \ref{tab:fdd_duration_baseline}, the regression analysis examining the effect of different public transport use durations on First Fixation Duration (FFD) in the \textit{Current version} environment did not reveal any statistically significant relationships. None of the tested duration categories (15--30 min, 30--45 min, 45--60 min, “Don’t know,” and more than one hour) produced a notable deviation in FFD compared to the reference condition (Estimate range: -14.100 to 24.256 ms, \(p > 0.05\)).] As with previous findings, these nonsignificant results may be influenced by limitations such as sample size or variability in the data.

\begin{table}[ht]
\centering
\caption{Regression Results for the effect of public transportation use duration on the first fixation duration in the ``Current version''}
\begin{tabular}{@{}lcccc@{}}
\toprule
\textbf{Variable}                              & \textbf{Estimate} & \textbf{Std. Error} & \textbf{t value} & \textbf{Pr(>|t|)} \\ \midrule
(Intercept)                                    & 272.500           & 20.458              & 13.320           &  $<$2e-16 ***     \\
15--30 min                     & -14.100           & 26.412              & -0.534           & 0.594             \\
30--45 min                     & 15.763            & 27.367              & 0.576            & 0.565             \\
45--60 min                     & -11.981           & 29.725              & -0.403           & 0.687             \\
Don’t know                     & 2.554             & 24.253              & 0.105            & 0.916             \\
More than one hour             & 24.256            & 26.922              & 0.901            & 0.368             \\ \midrule

\end{tabular}
\label{tab:fdd_duration_baseline}
\end{table}

\paragraph{Poorly Maintained}

\noindent As shown in Table~\ref{tab:ffd_duration_trash}, the regression analysis examining the effect of varying public transport use durations on First Fixation Duration (FFD) in the \textit{Poorly maintained current version} environment did not identify any statistically significant relationships for the duration variable. Although the “More than half an hour” category had an estimate of 8.43~ms, this difference was not significant \((p = 0.62043)\). While no effect of duration was detected, it is noteworthy that identifying as Non-white was associated with a statistically significant reduction in FFD \((p = 0.00593)\). These results should be interpreted cautiously, given potential limitations such as sample size or data variability.

\begin{table}[ht]
\centering
\caption{Regression Results for the effect of public transportation use duration on the first fixation duration in the ``Poorly maintained current version''}
\begin{tabularx}{\textwidth}{@{}Xcccc@{}}
\toprule
\textbf{Variable} & \textbf{Estimate} & \textbf{Std. Error} & \textbf{t value} & \textbf{Pr(>|t|)} \\
\midrule
Intercept  & 280.24 & 15.34 & 18.269 & <2e-16*** \\
Duration: More than half an hour  & 8.43   & 16.99 & 0.496  & 0.62043 \\
Gender: Non-male                  & 17.60  & 16.99 & 1.036  & 0.30146 \\
Ethnicity: Non-white              & -50.17 & 18.02 & -2.783 & 0.00593** \\
\bottomrule
\end{tabularx}
\label{tab:ffd_duration_trash}
\end{table}


\paragraph{Enhanced Version}

\noindent As shown in Table~\ref{tab:ffd_duration_enhanced}, when using a more aggregated set of public transport use duration categories, no statistically significant relationships emerged between duration and First Fixation Duration (FFD). In preliminary analyses using more finely grained categories as described in the Methodology section, certain durations appeared to be significant predictors. However, after combining categories for a more robust approach, these effects diminished, suggesting that the initial significance may have been sensitive to how duration was defined. Nonetheless, identifying as Non-white remained a significant predictor, associated with a 41.66~ms reduction in FFD \((p = 0.0147)\). As before, these results should be interpreted with caution, considering that changes in categorization, sample size, or other sources of variability can influence observed effects.

\begin{table}[ht]
\centering
\caption{Regression Results for the effect of public transportation use duration on the first fixation duration in the ``Enhanced'' cabin design)}
\begin{tabularx}{\textwidth}{@{}Xcccc@{}}
\toprule
\textbf{Variable} & \textbf{Estimate} & \textbf{Std. Error} & \textbf{t value} & \textbf{Pr(>|t|)} \\
\midrule
Intercept & 304.127 & 14.395 & 21.127 & <2e-16*** \\
Duration: More than half an hour  & 4.099   & 15.948 & 0.257  & 0.7975 \\
Gender: Non-male                  & 1.140   & 15.943 & 0.071  & 0.9431 \\
Ethnicity: Non-white              & -41.663 & 16.916 & -2.463 & 0.0147* \\
\bottomrule
\end{tabularx}
\label{tab:ffd_duration_enhanced}
\end{table}

\paragraph{Biophilic Design}

\noindent As shown in Table~\ref{tab:ffd_duration_plants}, when using a more aggregated set of public transport use duration categories for the \textit{Biophilic} cabin design, none of the duration categories emerged as significant predictors of First Fixation Duration (FFD). In earlier, more detailed analyses (not presented here), certain durations appeared to significantly influence FFD; however, after consolidating categories into a broader classification, those effects dissipated. This suggests that the observed significance was sensitive to how the duration variable was defined. As before, these results should be interpreted cautiously, considering that changes in variable categorization, sample size, or other sources of variability can affect the significance of observed effects.

\begin{table}[ht]
\centering
\caption{Regression Results for the effect of public transportation use duration on the first fixation duration in the ``Biophilic'' cabin design}
\begin{tabularx}{\textwidth}{@{}Xcccc@{}}
\toprule
\textbf{Variable} & \textbf{Estimate} & \textbf{Std. Error} & \textbf{t value} & \textbf{Pr(>|t|)} \\
\midrule
Intercept  & 283.877 & 14.215 & 19.971 & <2e-16*** \\
Duration: More than half an hour  & 12.509  & 15.749 & 0.794  & 0.428 \\
Gender: Non-male                  & 16.881  & 15.744 & 1.072  & 0.285 \\
Ethnicity: Non-white              & -3.769  & 16.704 & -0.226 & 0.822 \\
\bottomrule
\end{tabularx}
\label{tab:ffd_duration_plants}
\end{table}

\paragraph{Bike}
\noindent As shown in Table~\ref{tab:ffd_duration_bike}, the regression analysis for the \textit{Bike-centered} cabin design did not reveal any statistically significant relationships between public transport use duration and First Fixation Duration (FFD). None of the tested duration categories produced notable deviations from the baseline (all \(p > 0.05\)). As with other nonsignificant findings, these results should be interpreted with caution, acknowledging potential limitations such as sample size or variability in the data.

\begin{table}[ht]
\centering
\caption{Regression Results for the effect of public transportation use duration on the first fixation duration in the ``Bike-centered'' cabin design}
\begin{tabularx}{\textwidth}{@{}Xcccc@{}}
\toprule
\textbf{Variable} & \textbf{Estimate} & \textbf{Std. Error} & \textbf{t value} & \textbf{Pr(>|t|)} \\
\midrule
Intercept  & 294.935 & 14.331 & 20.580 & <2e-16*** \\
Duration: More than half an hour  & 15.016  & 15.878 & 0.946  & 0.3455 \\
Gender: Non-male                  & -6.671  & 15.873 & -0.420 & 0.6747 \\
Ethnicity: Non-white              & -29.427 & 16.841 & -1.747 & 0.0822. \\
\bottomrule
\end{tabularx}
\label{tab:ffd_duration_bike}
\end{table}

\paragraph{Productivity}

\noindent As shown in Table~\ref{tab:ffd_duration_productivty}, the regression analysis for the \textit{Productivity} cabin design did not reveal any statistically significant effects of public transport use duration on First Fixation Duration (FFD). While estimates varied across categories, none approached conventional significance thresholds \((p > 0.05)\). As with the other nonsignificant findings, these results should be interpreted with caution, acknowledging potential limitations such as sample size or variability in the data.

\begin{table}[ht]
\centering
\caption{Regression Results for the effect of public transportation use duration on the first fixation duration in the ``Productivity'' cabin design}
\begin{tabularx}{\textwidth}{@{}Xcccc@{}}
\toprule
\textbf{Variable} & \textbf{Estimate} & \textbf{Std. Error} & \textbf{t value} & \textbf{Pr(>|t|)} \\
\midrule
Intercept  & 276.90 & 14.34 & 19.307 & <2e-16*** \\
Duration: More than half an hour  & 24.21  & 15.89 & 1.524  & 0.129 \\
Gender: Non-male                  & 15.73  & 15.88 & 0.990  & 0.323 \\
Ethnicity: Non-white              & -26.67 & 16.85 & -1.582 & 0.115 \\
\bottomrule
\end{tabularx}
\label{tab:ffd_duration_productivty}
\end{table}

\end{appendices}

\end{document}